\documentclass[aps, prl, twocolumn, groupedaddress, superscriptaddress, shortbibliography, notitlepage, footinbib]{revtex4-1}

\usepackage{graphicx} 
\usepackage{color} 
\usepackage{xcolor}
\usepackage{amsmath}
\usepackage{amsfonts} 
\usepackage{amssymb} 
\usepackage{dcolumn}
\usepackage[justification=centerlast]{caption}
\usepackage{enumitem}
\usepackage{xr}
\usepackage{hyperref}
\hypersetup{colorlinks=true, urlcolor=blue,
  linkcolor=blue, citecolor=blue}
\usepackage{subcaption}
\usepackage{stackengine}
\newcommand{\CID}{\mathcal{K}}

\usepackage{wasysym}

\definecolor{mygray}{gray}{0.8}
\definecolor{plum}{rgb}{.5,0,1}
\begin{document}
\title{Correlation lengths in the language of computable information}
\author{Stefano Martiniani}
\email{mart5523@umn.edu}
\affiliation{Department of Chemical Engineering and Materials Science, University of Minnesota, Minneapolis, Minnesota 55455, USA}
\affiliation{Center for Soft Matter Research, Department of Physics, New York University, New York 10003, USA}
\author{Yuval Lemberg}
\affiliation{Department of Physics, Technion - IIT, 32000 Haifa, Israel}
\author{Paul M. Chaikin} 
\email{chaikin@nyu.edu}
\affiliation{Center for Soft Matter Research, Department of Physics, New York University, New York 10003, USA}
\author{Dov Levine} 
\email{dovlevine19@gmail.com}
\affiliation{Department of Physics, Technion - IIT, 32000 Haifa, Israel}

\begin{abstract}
Computable Information Density (CID), the ratio of the length of a losslessly compressed data file to that of the uncompressed file, is a measure of order and correlation in both equilibrium and nonequilibrium systems. Here we show that correlation lengths can be obtained by decimation, thinning a configuration by sampling data at increasing intervals and recalculating the CID. When the sampling interval is larger than the system's correlation length, the data becomes incompressible. The correlation length and its critical exponents are thus accessible with no \textit{a-priori} knowledge of an order parameter or even the nature of the ordering. The correlation length measured in this way agrees well with that computed from the decay of two-point correlation functions $g_{2}(r)$ when they exist. But the CID reveals the correlation length and its scaling even when $g_{2}(r)$ has no structure, as we demonstrate by ``cloaking'' the data with a Rudin-Shapiro sequence.
\end{abstract}

\maketitle

Physics, and indeed science in general, is a search to find and quantify correlations and order in nature. In many cases this organization is evident and quantifiable in terms of an order parameter which is identified with a broken symmetry. Such symmetry breaking is often associated with a phase transition at which the order parameter becomes finite, and a length scale for the persistence of the order which diverges as one approaches the transition. There are, however, systems in nature whose order we do not yet understand or for which we cannot define an order parameter in the conventional sense. Even in such cases, we may reasonably expect that there exist some as yet unidentified correlations, with associated length scales which may or may not diverge. 

The basic idea we wish to exploit is the intimate connection between order and information: it takes less information to completely describe a system with correlations than an uncorrelated one. The basis for the quantification of these ideas can be found in information theory \cite{cover2012elements}, in particular the Shannon entropy \cite{shannon2001mathematical} and the Kolmogorov complexity \cite{kolmogorov1968three,chaitin1966length}. In recent work \cite{martiniani2019quantifying} we have introduced a quantitative measure, the Computable Information Density (CID), $\CID \equiv \mathcal{L}(\mathbf{x})/L$, that is the binary code length, $\mathcal{L}(\mathbf{x})$, of a losslessly compressed file $\mathbf{x}$ (such as the microstate of a many-body system) divided by the uncompressed length $L$ (the number of degrees of freedom) of $\mathbf{x}$ \footnote{Notice that the CID is not the same as the compression ratio (or compressibility) $\varrho$ of the sequence, in fact $\text{CID}=\varrho \log_2 |\alpha|$, where $|\alpha|$ is the dictionary size of the sequence \cite{ziv1978compression}.}, which is closely related to the Shannon and Kolmogorov measures, and which is an excellent approximant of the thermodynamic entropy, $S$, for equilibrium systems. In what follows we estimate the CID using the unrestricted Lempel-Ziv string matching algorithm (LZ77) \cite{ziv1977universal, shields1999performance}, a \textit{universal} (\textit{i.e.,} requires no \textit{a-priori} knowledge of the nature of the ensemble)  and asymptotically \textit{optimal} code (\textit{i.e.}, $\lim_{L \to \infty} \CID = S$) \cite{cover2012elements, martiniani2019quantifying}. CID reveals the nature of phase transitions (first or second order), the position of critical points, and the exponent of critical slowing down, for both equilibrium and nonequilibrium phase transitions. Here we wish to explore whether CID can be used to determine correlation lengths for such systems \footnote{The use of information to find what might be regarded as a correlation length for written English was the subject of Reference \cite{shannon1951prediction} in the context of predicting letters in a string, where correlations of up to eight letters were inferred. A similar idea was discussed in Reference \cite{kurchan2010order} where patch entropy, calculated with different block sizes, was employed to discuss correlation lengths in physical systems.}.

The standard method for computing the correlation length $\xi$ of a system is to calculate some correlation function, typically two-point, and see how it decays with distance. This presupposes that the order and proper correlation function is known. In this paper, we propose a method that does not require this knowledge, which is based on the fundamental idea that correlations reduce the CID of a system. 

If a system consists of uncorrelated elements, the CID takes its maximum value. To exploit this, we sample a system on various length scales $\Delta$ by culling out degrees of freedom on smaller scales. In Fig.~1a-b we consider a 1D model of randomly placed hard rods of length $\ell=4$, while in Fig.~1c-d we have a 1D Ising model \cite{ising1925beitrag} at finite temperature and zero applied magnetic field. The diagrams in Fig.~1a and 1c show a respective configuration from each of the models, which we sample on every fourth site ($\Delta=4$). If $\Delta < \xi$, the remaining degrees of freedom still show correlations, albeit weakened, but if $\Delta > \xi$, all correlations are lost and the CID attains its maximal value. In the simplest cases \textit{e.g.} for the 1D hard-rod model, to estimate $\xi$ we can simply look for the smallest value of $\Delta$ where the CID reaches its maximum. However, this is not always adequate \textit{e.g.} in the 1D Ising model the CID approaches its maximum exponentially, so we find that in general it is better to study the way that the CID scales with $\Delta$ by collapsing the data. This procedure has the advantage of being independent of the system being analyzed.

\begin{figure*}
\centering
\begin{subfigure}{.5\textwidth}
\topinset{\bfseries(b)}{\topinset{\bfseries(a)}{\includegraphics[width=1\linewidth]{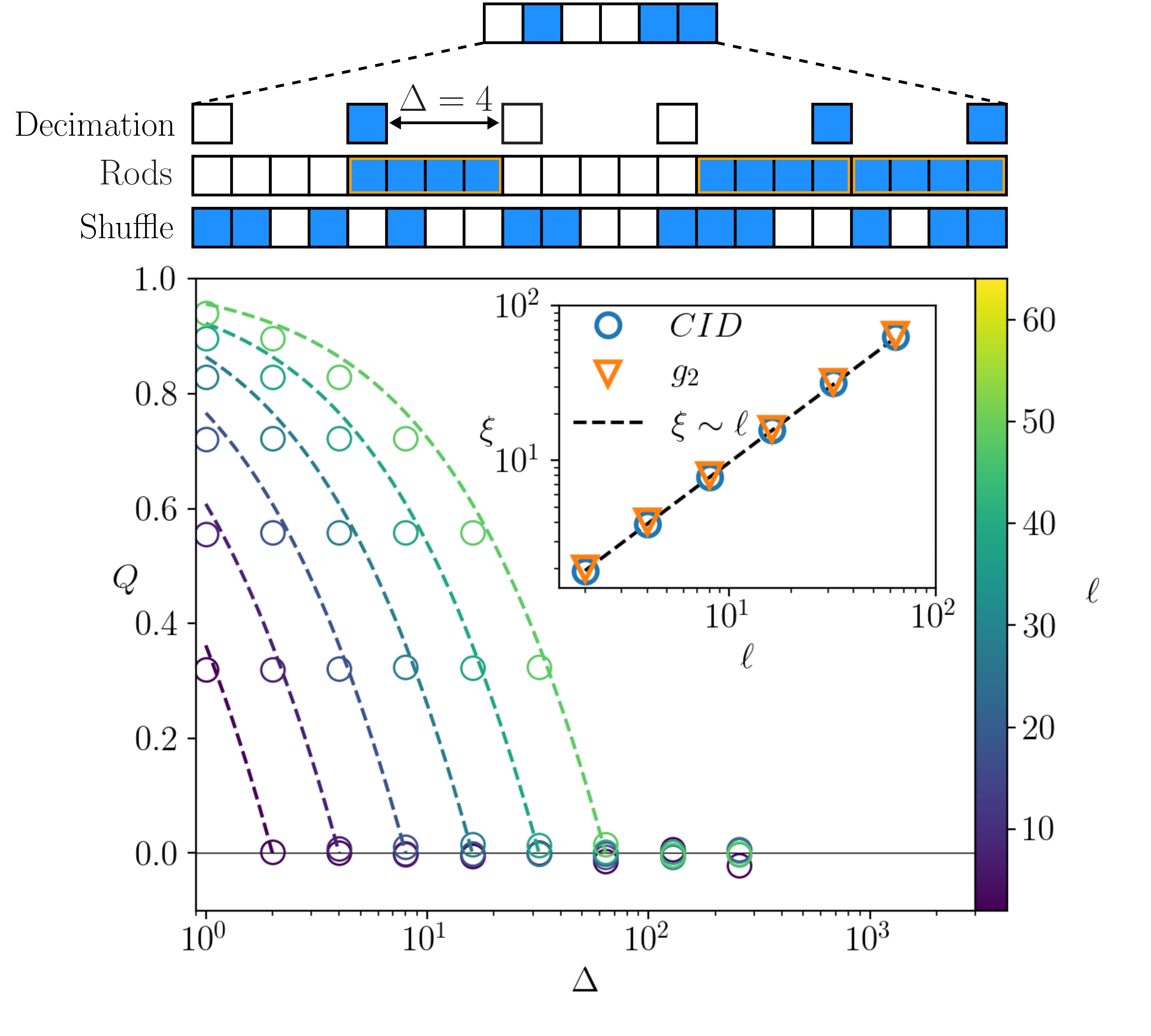}}{0in}{-1.6in}}{0.8in}{-1.6in}
\end{subfigure}%
\begin{subfigure}{.5\textwidth}
\topinset{\bfseries(d)}{\topinset{\bfseries(c)}{\includegraphics[width=0.94\linewidth]{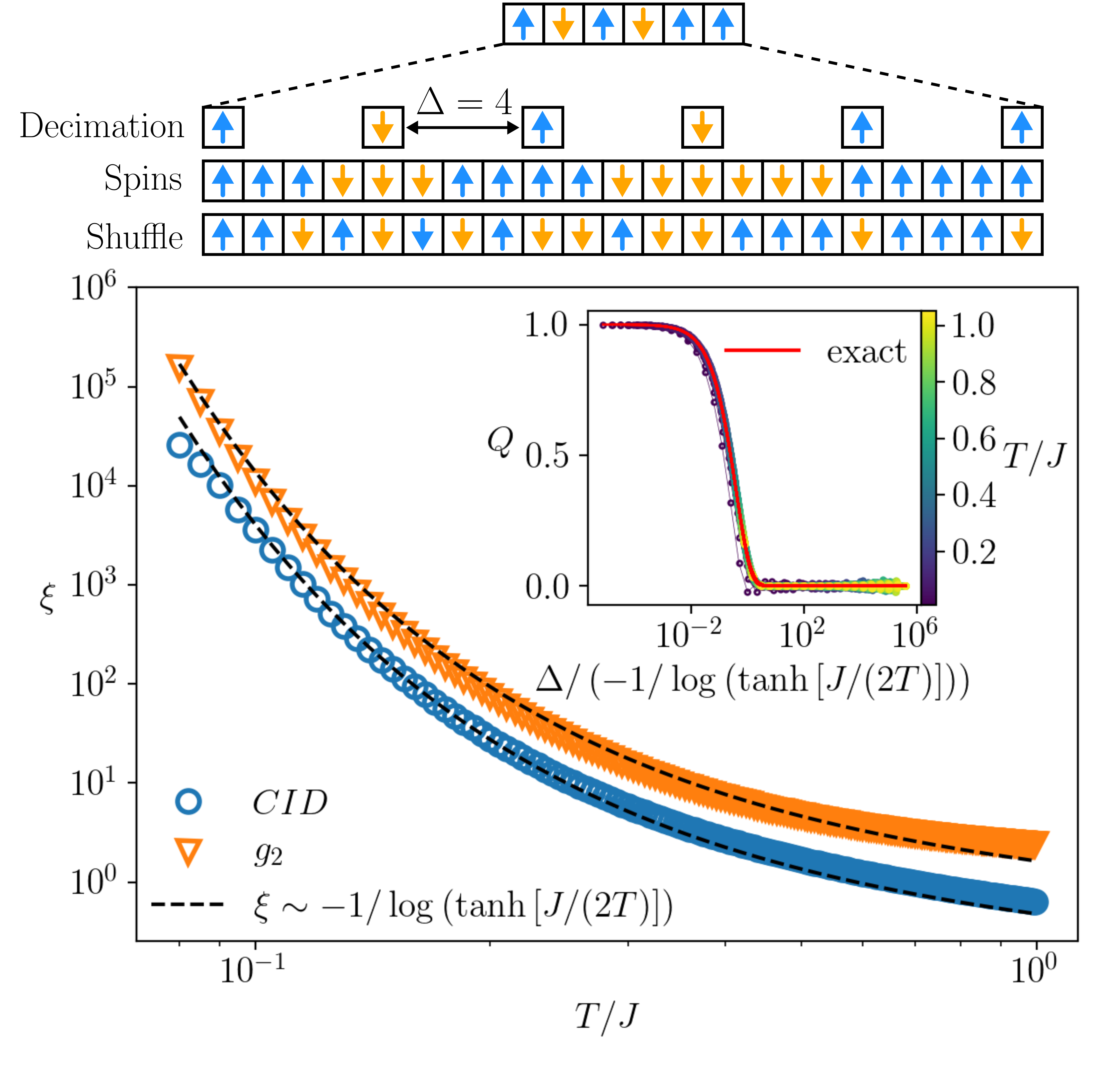}}{0in}{-1.6in}}{0.8in}{-1.6in}
\end{subfigure}
\protect\caption{Diagrams depict (a) 1D hard-rods of length $\ell = 4$ and (c) 1D Ising configurations, their shuffled counterparts, and the result of decimation when the sampling interval is $\Delta=4$.  (b) 1D hard rods, with lengths $\ell=2^{i}$ ($1 \leq i \leq 6$), randomly distributed on a grid of length $L = 2^{16+i}$, at fixed density $\rho \equiv N_r \ell / L = 1/4$.  Upon decimation at intervals $\Delta > \ell$, the configurations reduce to a random sequence. The main panel shows $Q(\Delta,\ell)$, dashed lines are exact solution to leading order (see Eq.~S5 in SM \cite{Note4}). Inset shows $\xi$ as computed from $Q(\Delta)$ and $g_2(r)$ taking $\xi$ to be the value where $Q(\xi) = 0.025$ and where $g_2(\xi)$ is a its minimum (see SM \cite{Note4}). (d) 1D Ising model of size $L=2^{20}$ simulated by Wolff algorithm \cite{wolff1989collective}. We performed the same analysis as for panel $a$ but extracted $\xi$ by fitting the curves to exponential functions of the form $g_2(r) = g_2(0)\exp(-r/\xi)$ and ${Q}(\Delta) = {Q}(1)\exp(-(\Delta-1)/\xi)$. Data were averaged over 200 equilibrium configurations. The black dashed lines show the theoretical scaling for $\xi$ with T. Inset shows the collapsed $Q(\Delta,T)$, along with the analytical solution Eq.~\ref{eq:Fdef} (red line). \label{fig:1}} 
\end{figure*}

In particular, we wish to study the quantity \footnote{In principle, Q depends on control parameters such as temperature or density which are suppressed here; they will be indicated where relevant. }
\begin{equation}
{Q}(\Delta) \equiv 1- \frac{\CID (\Delta)} {\CID_{\mathrm{shuf}}(\Delta)}
\label{eq:Fdef}
\end{equation}
where we denoted the CID as $\CID$, and the subscript `shuf' refers to a configuration obtained by randomly shuffling all its degrees of freedom. Because a randomly shuffled configuration has no correlations, $\CID_{\mathrm{shuf}}(\Delta) \geq \CID(\Delta)$, and $0 \leq {Q} \leq 1$. 

The 1D hard-rod system consists of $N_r$ rods, each occupying $\ell$ contiguous sites, randomly distributed on a lattice of length $L$ sites. The fraction of occupied sites is $\rho=N_r \ell /L$. Configurations of the system are represented by strings $\{n_{j}\}$, where $j = 1, 2, \ldots, L$ and $n_j = 1$ if site $j$ is occupied by a rod element, and $n_j=0$ if it is not. The trivial correlation length is $\ell$. Can we discover this by decimating configurations, computing their CID, and estimating $\xi_{\mathrm{CID}}$ from the value of $\Delta$ at which ${Q}(\Delta,\ell) \rightarrow 0$ for different values of $\ell$? 

The decimated configurations are obtained by retaining the occupancies $n_{j \cdot \Delta}$ (with $j=1,2,\ldots, L/\Delta$) of a configuration, deleting all the others, and then rescaling the system by a factor $\Delta$. In Fig.~1b we plot $Q(\Delta)$ for several values of $\ell$, with $\xi_{\mathrm{CID}}$ \textit{vs.} $\ell$ shown in the inset, along with the values of $\xi_{g_2}$ obtained by finding the minimum of the two-point correlation function $g_2(r)$ of the undecimated configurations (see SM \footnote{See Supplementary Material [url] for a definition of the models, exact solutions, supplementary data and implementation details, which include Refs. \cite{onsager1944crystal, codello2015approximating, binney1992theory, golay1949multi, golay1951static, rudin1959some, constantinescu2007lempel, sweetsourcod, karkkainen2013linear, lz77alg, skilling2004programming, hilbertcurve}.}). Both correlation lengths are close in value to $\ell$ but differ numerically by a small factor. The data collapse, along with the exact result, showing that ${Q}(\rho,\Delta,\ell) \rightarrow 0$ linearly with $\Delta-\ell$ as $\Delta \rightarrow \ell$, are given in SM \cite{Note4}.

We next consider the equilibrium 1D Ising model, which has a transition at $T=0$. Both the entropy $S$ and $\xi$ may be solved for exactly \cite{salinas2001introduction}, and give \footnote{Where we use the analytic value for the entropy in place of the CID in Equation \ref{eq:Fdef}. See SM \cite{Note4} for a full derivation.} 
\begin{equation}
{Q}(\Delta, \xi) = \frac{e^{-\Delta /\xi } \coth
   ^{-1}\left(e^{\Delta /\xi }\right)+\frac{1}{2} \log
   \left(1-e^{-2
   \Delta /\xi }\right)}{\log (2)}
   \label{eq:q1d}
\end{equation}
Here, ${Q} \rightarrow 0$ exponentially as ${Q}\sim e^{-2 \Delta / \xi}$, making an extrapolation inadequate to determine $\xi_{\mathrm{CID}}$. Rather, we generate equilibrium spin configurations for different temperatures, decimate these configurations, calculate the CID to find ${Q}(\Delta)$, and then collapse them to a universal curve (inset of Fig.~1d). The collapse indicates that there is a single length scale $\xi$ in the problem and yields its temperature dependence. An exponential fit to a single curve ${Q}$ \textit{vs.} $\Delta/\xi$ then gives the value of $\xi_{\text{CID}}$. $\xi_{\text{CID}}(T)$ and $\xi_{g_{2}}(T)$ are shown in Fig.~1d; both show the same $T$ dependence as the analytic $\xi(T)$. 

\begin{figure}
\centering
\includegraphics[width=\columnwidth]{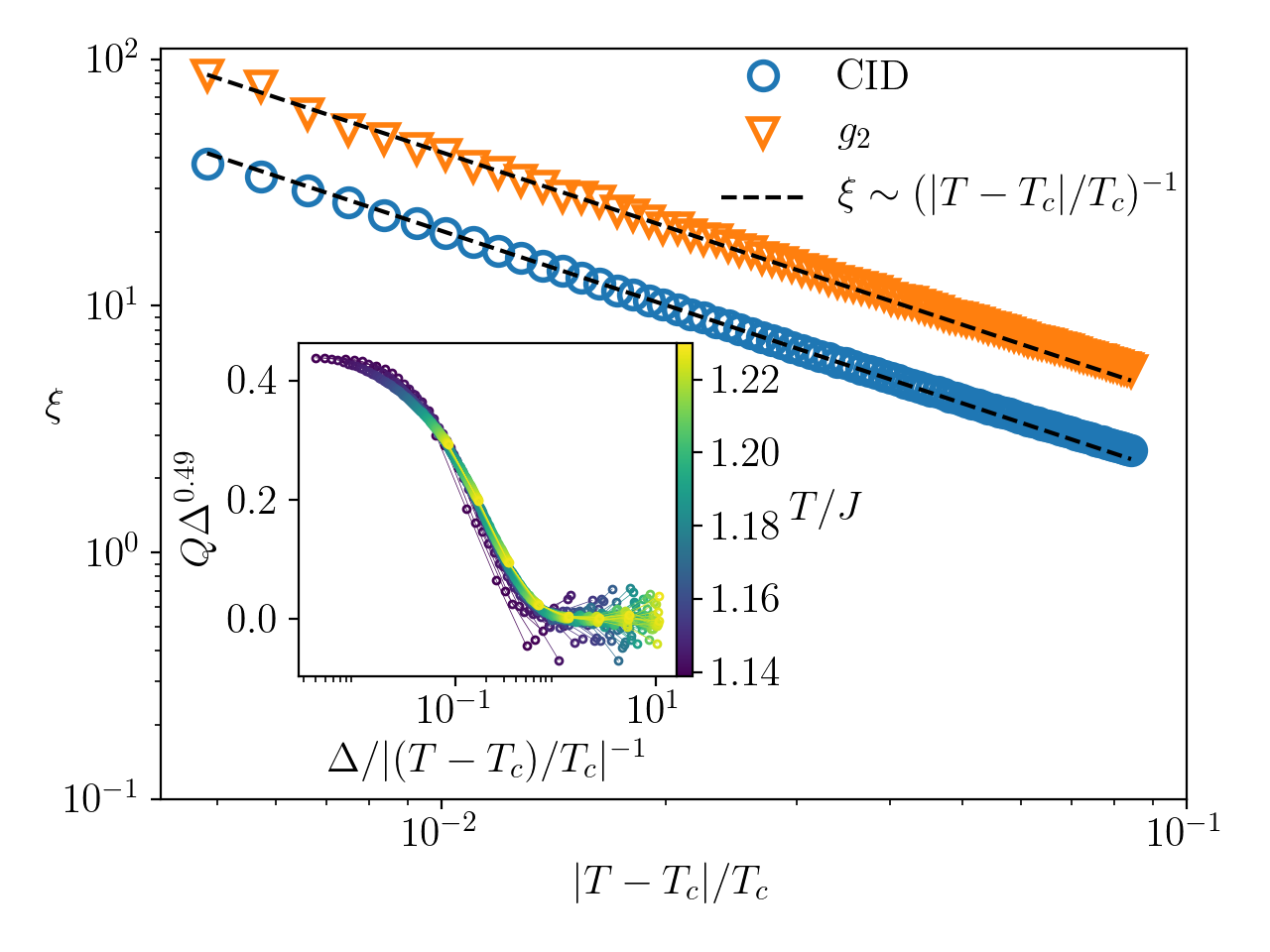}
\protect\caption{2D Ising model of size $L=2^{10} \times 2^{10}$ simulated by Wolff algorithm \cite{wolff1989collective}. Correlation lengths $\xi$ extracted by fitting the pair-correlation function to $g_2(r) = g_2(0)\exp(-r/\xi)/r^\eta$ and ${Q}(\Delta) = {Q}(1)\exp(-(\Delta-1)/\xi)/\Delta^\theta$. We find $\eta \approx 1/4$ and $\theta \approx 1/2$. The black dashed lines show the theoretical scaling $\xi \sim |T-T_c|/T_c$ with $T_c \approx 1.1345$. Data were averaged over 200 equilibrium configurations. Inset shows the collapse of the scaled $Q(\Delta,T)$.} 
\end{figure}

Results in 2D for the $q$-state Potts models ($2\leq q \leq 8$) \cite{potts1952some, wu1982potts} are shown in the SM \cite{Note4}. For $q=2$, this is the Ising model. Fig.~2 (inset) shows the collapse of ${Q}$ obtained by scaling the axes; this allows us to determine the critical exponent to be $\nu = 1$, where $\xi(T) \propto (T-T_c)^{-\nu}$. Fitting $Q(\Delta,T)$ at a single temperature gives us the numerical value of $\xi_{\mathrm{CID}}$, which is plotted alongside the value obtained from $g_{2}(r)$ in the main panel of Fig.~2. 
%In Fig.~2b we show a so-called \textit{image pyramid} \cite{burt1981fast, beyerer2015machine} obtained upon decimating an Ising configuration close to the critical temperature $T_c$. 
Notice that while decimating by $\Delta$ correctly yields configurations with correlation length $\xi/\Delta$, these are not equilibrium configurations with the same correlation length. This can be seen for instance from the fact that magnetization is invariant under decimation. In SM \cite{Note4} we consider an alternative blocking transformation known as \textit{majority rule} \cite{kadanoff1966scaling}, that yields valid equilibrium configurations and for which we can derive an exact expression for 2D Ising analogous to Eq.~\ref{eq:q1d}, and verify that there is good agreement between theory and numerical results in 2D.

We now consider the Conserved Lattice Gas (CLG), a dynamical nonequilibrium lattice model of the conserved directed percolation class \cite{hinrichsen_non-equilibrium_2000}. In the CLG (as illustrated in Fig.~3a) an occupied site is considered ``active'' if one of the nearest neighbors is also occupied (orange circles). Sites have a maximum occupancy of $1$ particle. At each time step, active sites are emptied stochastically by moving the particle to one of the empty neighboring sites (black arrows). The model has a continuous phase transition from a low density absorbing phase (where all sites are inactive) to a high density active phase where the dynamics persist forever. Configurations at the critical point are hyperuniform \cite{PRL_hyper, torquato_local_2003}. In 1D the critical density $\rho_{c} = 1/2$ corresponds to a periodic arrangement where every other site is occupied (\textit{i.e.}, $101010\dots$). In Fig.~3b we show $\xi(\rho)$ as obtained both from CID and $g_{2}(r)$, as well as the scaling collapse of $Q(\Delta, \rho)$ for different densities (inset). We find that $\xi(\rho)$ diverges with the exponent $\nu = 2$ for both measures as $\rho \to \rho_{c}$.

\begin{figure}[b]
\topinset{\bfseries(b)}{\topinset{\bfseries(a)}{\includegraphics[width=\columnwidth]{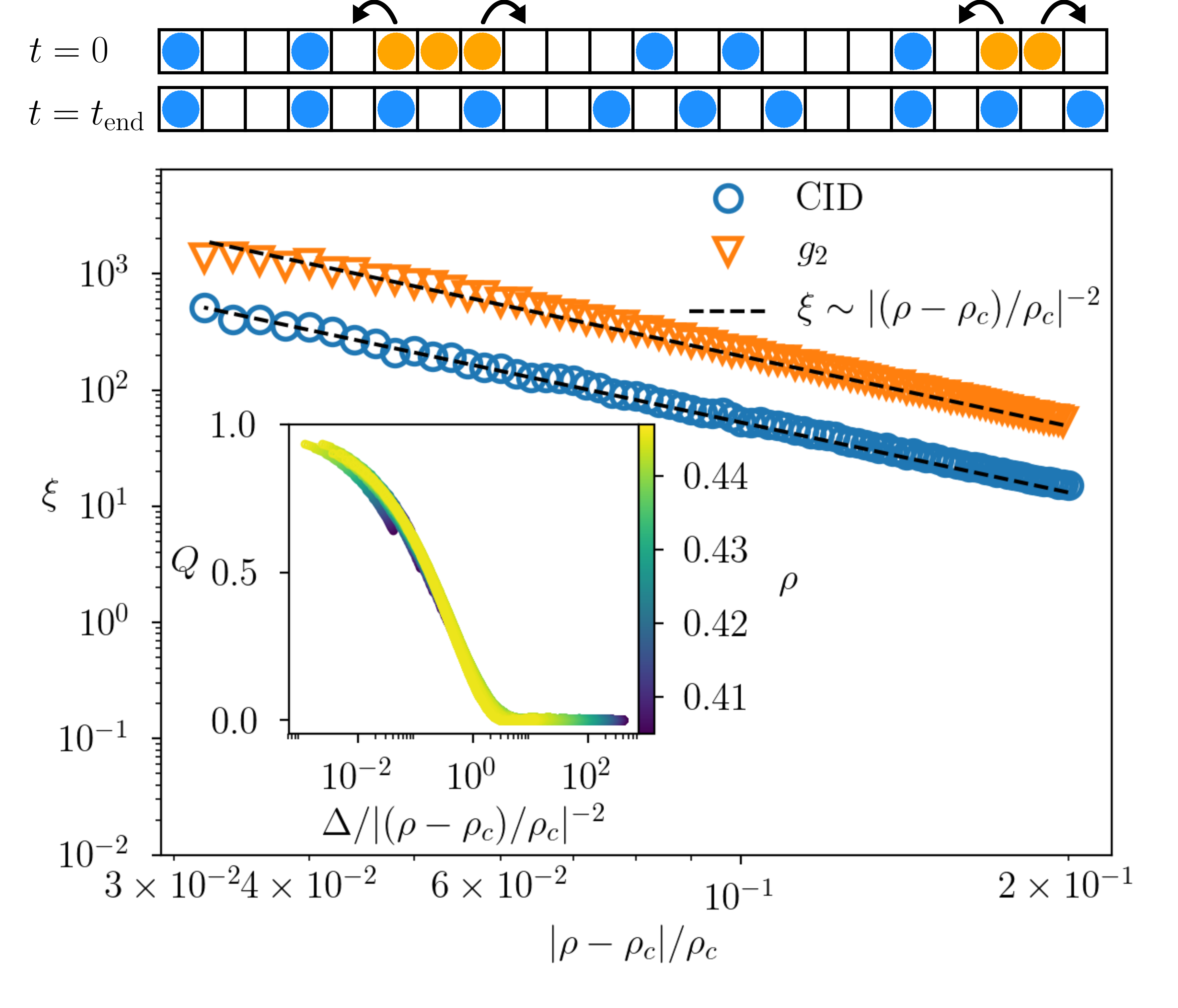}}{-0.1in}{-1.6in}}{0.57in}{-1.6in}
\protect\caption{1D Conserved Lattice Gas. (a) At time $t=0$ the system is in an active randomly sampled state (active sites in orange) and the possible moves prescribed by the dynamics are indicated by the arrows. When the density $\rho \leq \rho_c$ the system relaxes to an absorbing state with no active sites. (b) Correlation lengths $\xi$ for a system of size $L=2^{17}$ starting from randomly sampled states, extracted from $Q(\Delta)$ and $|g_2(r)|$ by taking $\xi$ to be the value where $Q(\xi) = 0.05$ and $|g_2(\xi)| = 0.05$. The black dashed lines show the fitted scaling $\xi \sim |(\rho - \rho_c)/\rho_c|^{-2}$, where $\rho_c = 0.5$. Inset shows the collapse of the scaled $Q(\Delta,\rho)$. Data were averaged over 15 independently generated configurations. \label{fig:3}}
\end{figure}

\begin{figure*}
\centering
\begin{subfigure}{.5\textwidth}
\topinset{\bfseries(a)}{\includegraphics[width=\linewidth]{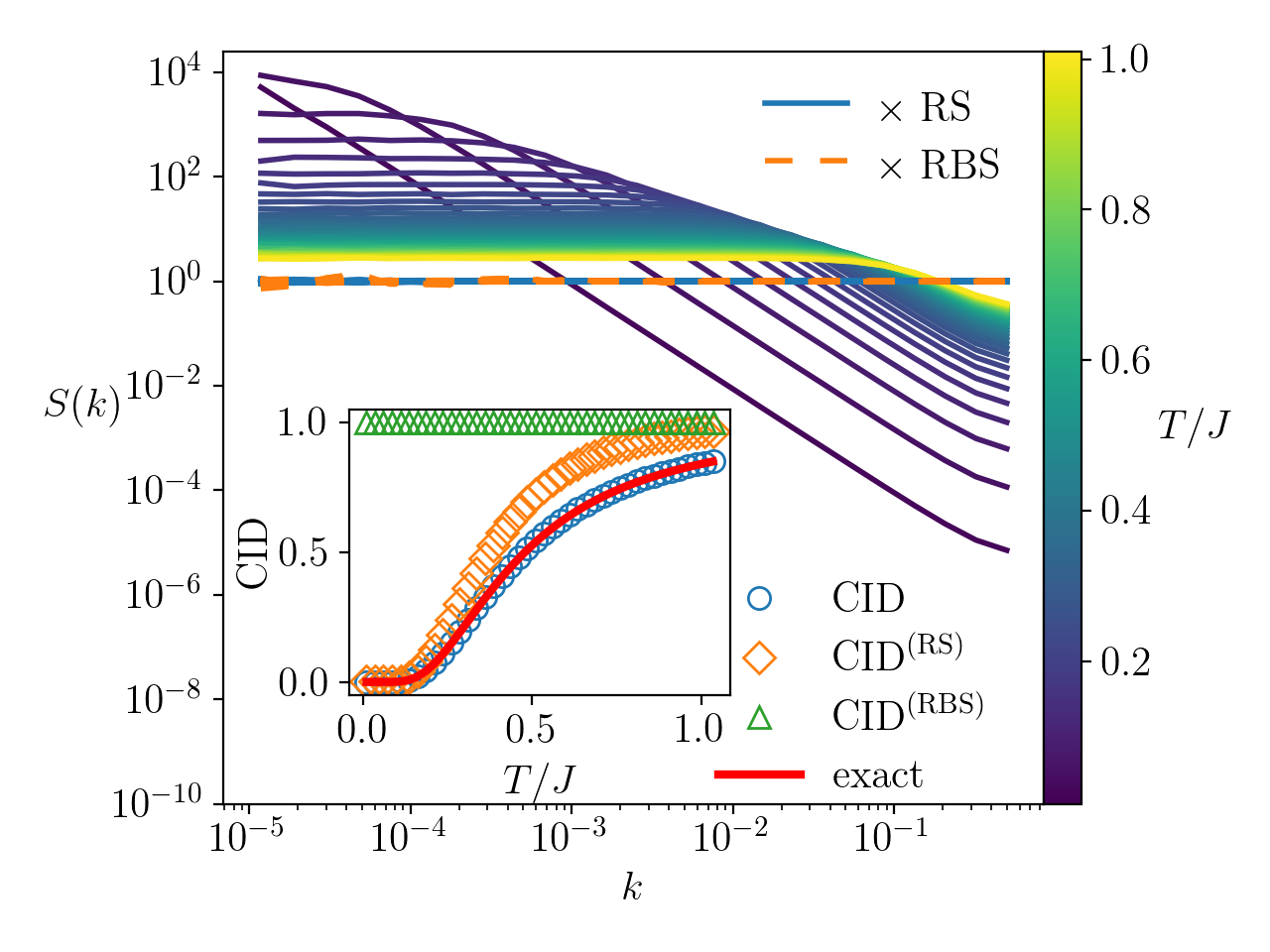}}{-0.05in}{-1.5in}
\end{subfigure}%
\begin{subfigure}{.5\textwidth}
\topinset{\bfseries(b)}{\includegraphics[width=\linewidth]{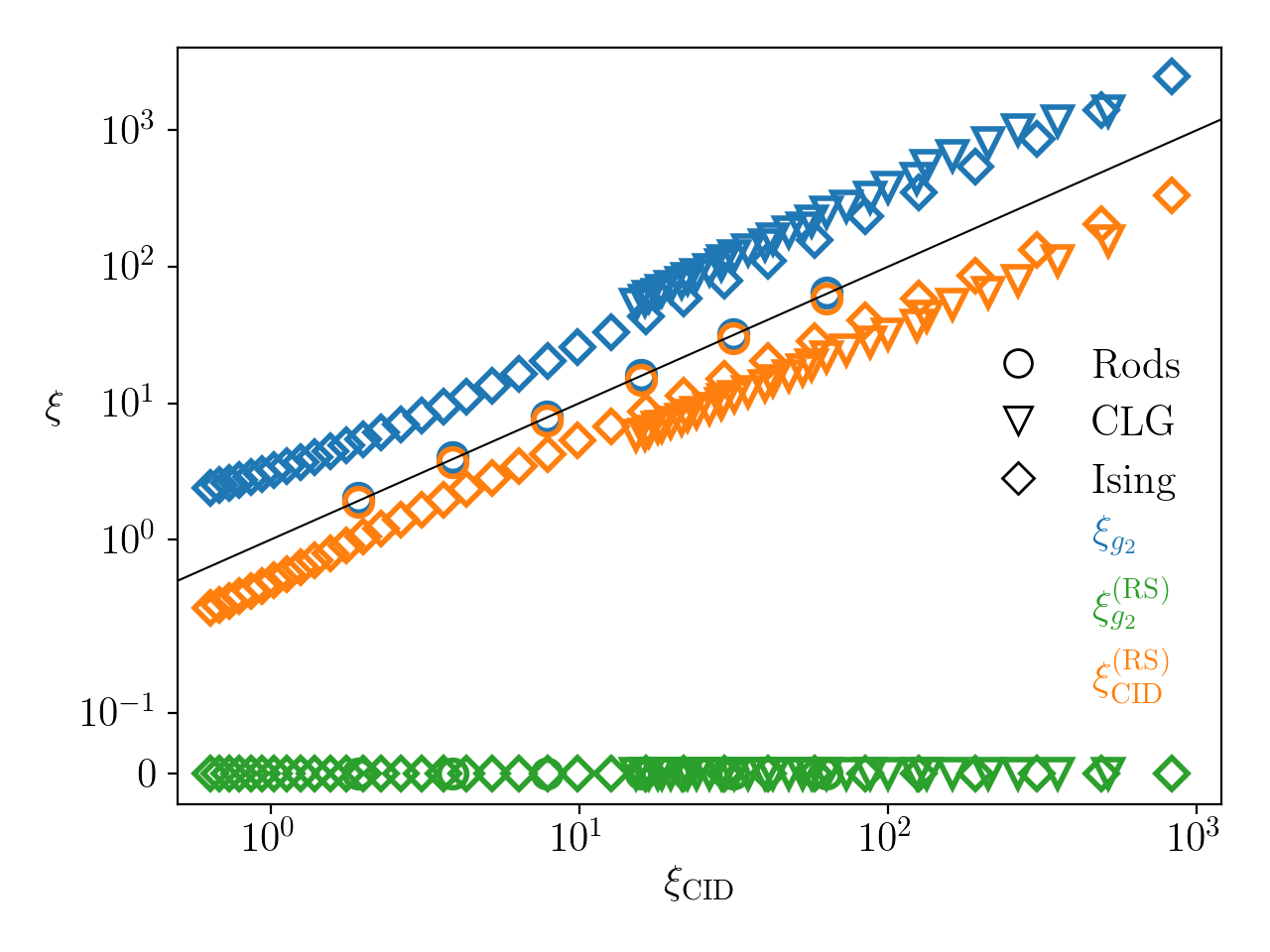}}{-0.05in}{-1.5in}
\end{subfigure}
\protect\caption{(a) Structure factor $S(k)$ for the 1D Ising model of size $L=2^{20}$. $S(k)$ is shown for the unaltered system, as well as for the Rudin-Shapiro, RS-cloaked, and Random Bernoulli Sequence (RBS) randomly-cloaked sequences, neither of which show any structure, indicating that all pair correlations have been destroyed. Inset: The CID as a function of T for the uncloaked and cloaked systems: The RS-cloaked system has nontrivial CID, but the randomly-cloaked system has maximum CID and is incompressible. Data were averaged over 200 equilibrium configurations. (b) Correlation lengths for the hard-rod, CLG, and 1D Ising systems, uncloaked (extracted from $g_2(r)$, in blue) and cloaked by RS (extracted from $g_2(r)$ and $Q(\Delta)$, in green and orange, respectively). The data is plotted against $\xi_{\mathrm{CID}}$, as measured by CID decimation of the uncloaked systems, so that a slope of 1 (solid line) means identical scaling for the RS-cloaked and uncloaked systems.
\label{fig:4}} 
\end{figure*}

We now want to see whether CID decimation can measure correlation lengths in systems with no two-point correlations. To this end, we will ``cloak'' strings in two ways that destroy their two-point correlations. To do this, we multiply 1D Ising configurations by (i) a random Bernoulli sequence (RBS) of equal numbers of $\pm 1$, and (ii) the deterministic Rudin-Shapiro sequence \cite{shapiro1952extremal, allouche2003automatic} (RS). Notice that this cloaking is exactly equivalent to studying two variants of the ``Mattis glass'', with RS and RBS ground states \cite{mattis1976solvable}. Both of these sequences have $g_2(j,k) = \delta_{jk}$, but while for RS the Kolmogorov complexity and CID tend to zero as the sequence length increases, they are maximal for RBS \cite{martiniani2019quantifying}.

Multiplying a sequence with structure in its $g_{2}(r)$ (or, equivalently, its structure factor $S(k)$) by RBS will produce a maximally random sequence with $g_{2}(r)=0$ for $r \ne 0$ (and $S(k)=1$ \footnote{There may also be a delta function at $q=0$.}). Moreover, this will increase its CID \footnote{Unless the sequence is itself already random, in which case there is no change.}, and cause all information about the original configuration to be lost (unless decoded by the identical random sequence). A similar multiplication by RS will remove all two-point correlations, also giving $g_{2}(r)=0$ (for $r \ne 0$) and $S(k)=1$, but will not appreciably change the Kolmogorov complexity of the original, since the RS itself has negligible Kolmogorov complexity. In this sense, ``cloaking'' by RS makes the sequence look random as far as $g_{2}(r)$ and $S(k)$ are concerned, while still retaining all the original information and order, although in a different form. We therefore expect that we should be able to recover the correlation length of the original, uncloaked system. 

In Fig.~4a (inset) we show the CID for 1D Ising configurations at different temperatures, and for the same configurations cloaked by RS and by RBS. 
RBS-cloaked 1D Ising has a flat $\text{CID} = 1$ indicating a correlation-less, maximally disordered system, but RS-cloaked 1D Ising retains much of its correlations. 
In the main panel of Fig.~4a we graph $S(k)$ for 1D Ising configurations, this shows increased correlations as T is lowered. Multiplying any configuration by RBS or RS  gives $S(k)$=1. We now perform the decimation procedure to determine the correlation length of the cloaked configurations. In Fig.~4b we plot the correlation lengths for the RS-cloaked configurations as determined from CID decimation and from $g_2(r)$, \textit{vs.} $\xi_{\mathrm{CID}}$ for the uncloaked 1D Ising configurations. For random cloaking, all information is lost, with no temperature dependence. However, for RS cloaking, although $\xi^{(RS)}_{g_{2}}(r)=0$, $\xi^{(RS)}_{\text{CID}}$ agrees well with the correlation length of the uncloaked 1D Ising system. Fig.~4b shows that similar conclusions hold also for the hard-rod and 1D CLG systems when cloaked by RS (see also SM \cite{Note4}). 

We note that RBS cloaking is analogous to the situation encountered in the analysis of static configurations of a square-lattice spin glass with random quenched disorder. In this case, without knowledge of the random couplings, the CID (or any other static estimator) would not be able to detect the order of an individual configuration because (like for RBS) individual configurations exhibit no correlations (and therefore are not compressible). It might, however, be possible to extract a correlation length by CID when considering the dynamics of the systems, \textit{viz.}  whole trajectories rather than individual configurations. We also argue that because structural glasses lack quenched disorder \cite{karmakar2013random}, the CID may be an effective tool for the analysis of the glass transition, \textit{e.g.} in soft sphere systems, given a sufficiently accurate CID estimator for continuum two and three dimensional systems.

CID decimation presents a simple and general method for finding the correlation length of equilibrium and nonequilibrium systems, or in fact of any temporal or spatial array (\textit{e.g.} a sequence or an image), with no {\it a-priori} knowledge of a possible order parameter, as well as in systems where two-point correlations are uninformative. We expect that this technique may lead to the discovery of order and aid in the quantification of correlation lengths in a wealth of new systems.

\begin{acknowledgments}
We particularly thank Yariv Kafri for crucial discussions at many stages of this project.
This work was primarily supported by the National Science Foundation Physics of Living Systems Grant 1504867. DL thanks the US-Israel Binational Science Foundation (grant 2014713), the Israel Science Foundation (grant 1866/16). P.M.C. was supported partially by the Materials Research Science and Engineering Center (MRSEC) Program of the National Science Foundation under Award DMR-1420073.
\end{acknowledgments}

%\subsection{Data availability}
%
%The datasets generated during the current study are available from  the corresponding author on reasonable request.
%
%\subsection{Code availability}
%
%CID was computed using the open-source library \textit{Sweetsourcod} \cite{sweetsourcod} wrapping the LZ77 factorization linear-time algorithms in Ref.~\cite{karkkainen2013linear, lz77alg}.
%
\bibliography{bibliography}
\end{document}

%
%\title{Measuring elusive correlation lengths}
\title{Supplementary Information: Correlation lengths in the language of computable information}
\author{Stefano Martiniani}
\email{mart5523@umn.edu}
\affiliation{Department of Chemical Engineering and Materials Science, University of Minnesota, Minneapolis, Minnesota 55455, USA}
\affiliation{Center for Soft Matter Research, Department of Physics, New York University, New York 10003, USA}
\author{Yuval Lemberg}
\affiliation{Department of Physics, Technion - IIT, 32000 Haifa, Israel}
\author{Paul M. Chaikin} 
\email{chaikin@nyu.edu}
\affiliation{Center for Soft Matter Research, Department of Physics, New York University, New York 10003, USA}
\author{Dov Levine} 
\email{dovlevine19@gmail.com}
\affiliation{Department of Physics, Technion - IIT, 32000 Haifa, Israel}

\maketitle
\beginsupplement
\section{Models}

\subsection{Ising}
The Ising model \cite{ising1925beitrag} is a simple model of ferromagnetism consisting of an array of up ($+1$) and down ($-1$) spins on a lattice, interacting through the Hamiltonian
\begin{equation}
\mathcal{H} = -J \sum_{\langle i, j \rangle} \sigma_i \sigma_j - B\sum_i \sigma_i
\end{equation}
where $J$ is the coupling constant, $\sigma_i = \pm 1$ is the \textit{i}-th spin, $B$ is the external magnetic field and $\langle i, j \rangle$ indicates all unique edges on the lattice. For $J > 0$ the system is ferromagnetic and undergoes spontaneous magnetization below the Curie temperature $T_c$. In this work we simulate 1- and 2-dimensional Ising models using the Hamiltonian of a $q=2$ Potts model, described next. 

\subsection{Potts}
The \textit{q}-state Potts model \cite{potts1952some, wu1982potts} is a generalization of the Ising model to more than 2 ($q$) spin orientations. The spins can take orientations
\begin{equation}
\theta_n = 2\pi n / q, ~~n = 0, 1, \dots, q-1
\end{equation}
and in the simplest form of the model (considered herein) for zero external field ($B=0$) the Hamiltonian takes the form
\begin{equation}
\mathcal{H} = - J \sum_{\langle i, j \rangle} \delta(n_i, n_j)
\end{equation}
where $\delta$ is the Kronecker delta. In this work we consider \textit{q}-state Potts models with $2 \leq q \leq 8$. For $q=2$ the Potts model is equivalent to the Ising model but with energies rescaled by a factor of $2$, so that in 2D: $T_c^{\text{(Potts)}} = T_c^{\text{(Ising)}}/2 \approx 1.1345$. \textit{q}-state Potts models on 2D square-lattices with periodic boundary conditions exhibit first order phase transitions for $q > 4$, and continuous transitions for $q \leq 4$. For $q$ close to the crossover values the transition becomes very weakly first order and it becomes difficult to distinguish from a continuous transition.

\subsection{Conserved Lattice Gas}
The Conserved Lattice Gas (CLG) in 1D is a simple model of nonequilibrium phase transition. Initially, N particles are distributed randomly on $L \geq N$ sites with no multiple occupancy. An occupied site is considered active if one of its neighbors is also occupied. The dynamics consist of moving particles randomly from active sites to unoccupied neighboring sites, as illustrated by the diagram in Fig.~3a (in practice we implement random sequential updates, so we displace one
particle at a time). The statistical state of the system is characterized by the order parameter $f_a$, the fraction of sites that are active. An ``absorbing state'' is attained when $f_a = 0$, at which point the dynamics end. No absorbing states are possible for densities higher than the geometrical limit $\rho_G = 0.5$. While for absorbing state models in general the critical density, $\rho_c$, does not coincide with $\rho_G$, the 1D CLG is atypical in that $\rho_c = \rho_G$. While above $\rho_c$ CLG behaves like an equilibrium system (all configurations satisfying the constraints introduced by the dynamics are accessible and equally probable), we previously showed \cite{martiniani2019quantifying} that this is not the case for $\rho < \rho_c$ where the system develops nontrivial correlations. In this work we show that for $\rho<\rho_c$ the correlation length grows with a nontrivial exponent $\nu = 2$, see Fig.~3b.

\section{Exact solutions}

\subsection{1D Hard rods}
The 1D equilibrium hard-rod system is amenable to trivial analytic solution. The degeneracy of the configurations, $\Omega$,  and that of a randomized system where the $N_r$ rods are broken up into $\ell$ particles each and randomly shuffled, $\Omega_{R}$, are:
\begin{equation}
\Omega = \frac{( N_r + L - N_r \ell )!}{N_r!\;(L-N_r\ell)!} ~~~~~~~~ \Omega_{R} = \frac{L!}{(N_r\ell)! \; (L-N_r\ell)!}
\end{equation}
where $L$ is the total number of sites and  $\rho \equiv N_r \ell / L$ is the density. We can explicitly calculate the normalized entropies, $S(\rho,\ell) = \log (\Omega) / L$ and $S_{R}(\rho,\ell) = \log(\Omega_{R})/L$ which are well approximated by the CID of single (large) configurations. We now rescale the system via $L \rightarrow L/\Delta$ and $\ell \rightarrow \ell/\Delta$, which corresponds to the  configuration obtained by decimating a fraction $1-1/\Delta$ of the sites, while leaving $\rho$ unchanged. The analytic result for $Q \equiv (S_{R}(\ell/\Delta) - S(\ell/\Delta))/S_{R}(\ell/\Delta)$ is given by using the normalized entropies instead of the CID.  To lowest order,
\begin{equation}
{Q}(\rho,\Delta, \ell) = \frac{(1-\rho+\rho \, \Delta/\ell)\log(1-\rho+\rho \, \Delta/\ell) + \rho \log \rho - (\rho  \, \Delta/\ell) \log(\rho \,  \Delta / \ell)}{-\rho \log \rho - (1-\rho) \log (1-\rho)}
\label{eq:F2}
\end{equation}

\begin{figure*}
\centering
\begin{subfigure}{.5\textwidth}
\topinset{\bfseries(a)}{\includegraphics[width=1\linewidth]{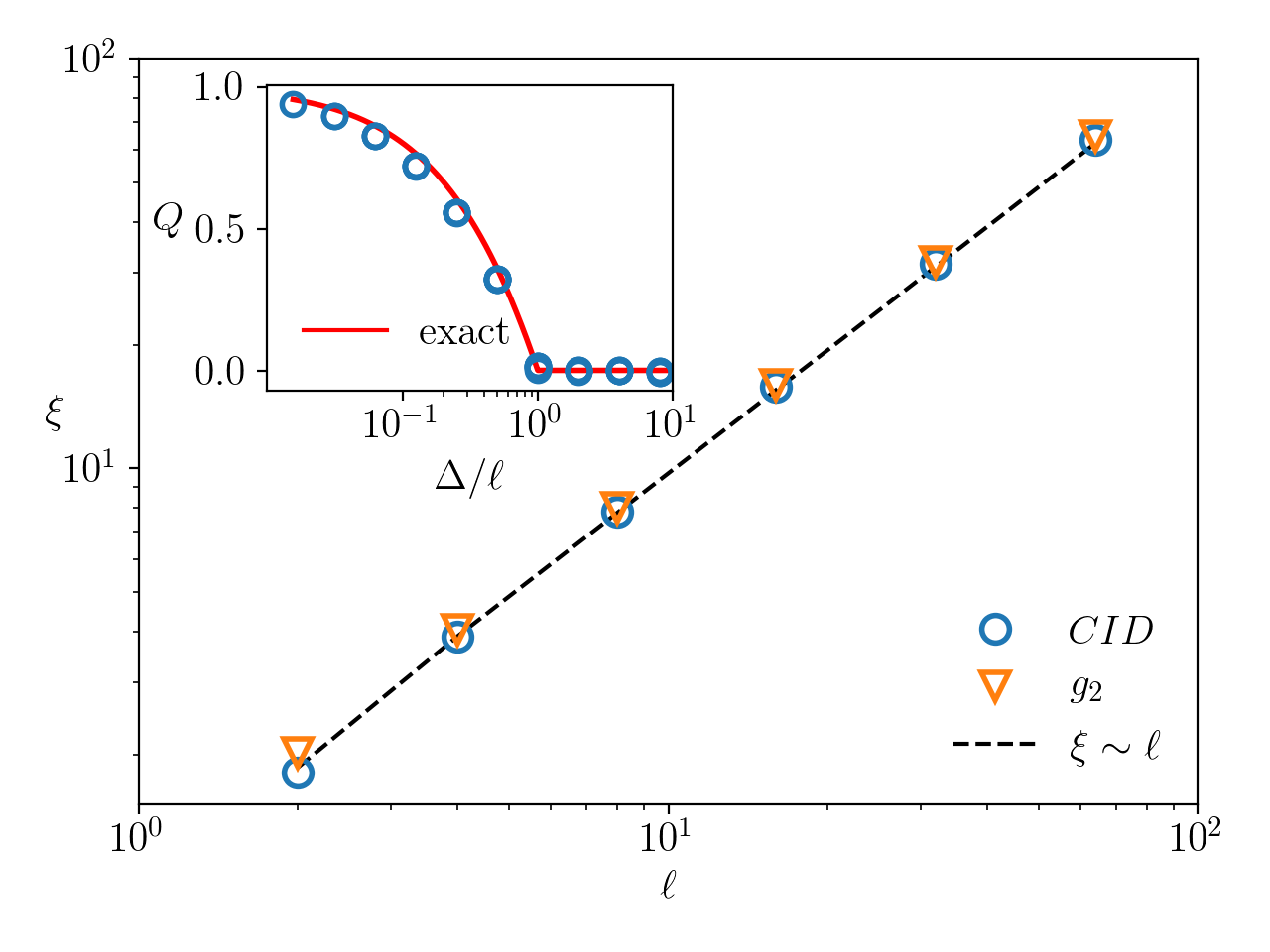}}{-0.05in}{-1.5in}
\end{subfigure}%
\begin{subfigure}{.5\textwidth}
\topinset{\bfseries(b)}{\includegraphics[width=1\linewidth]{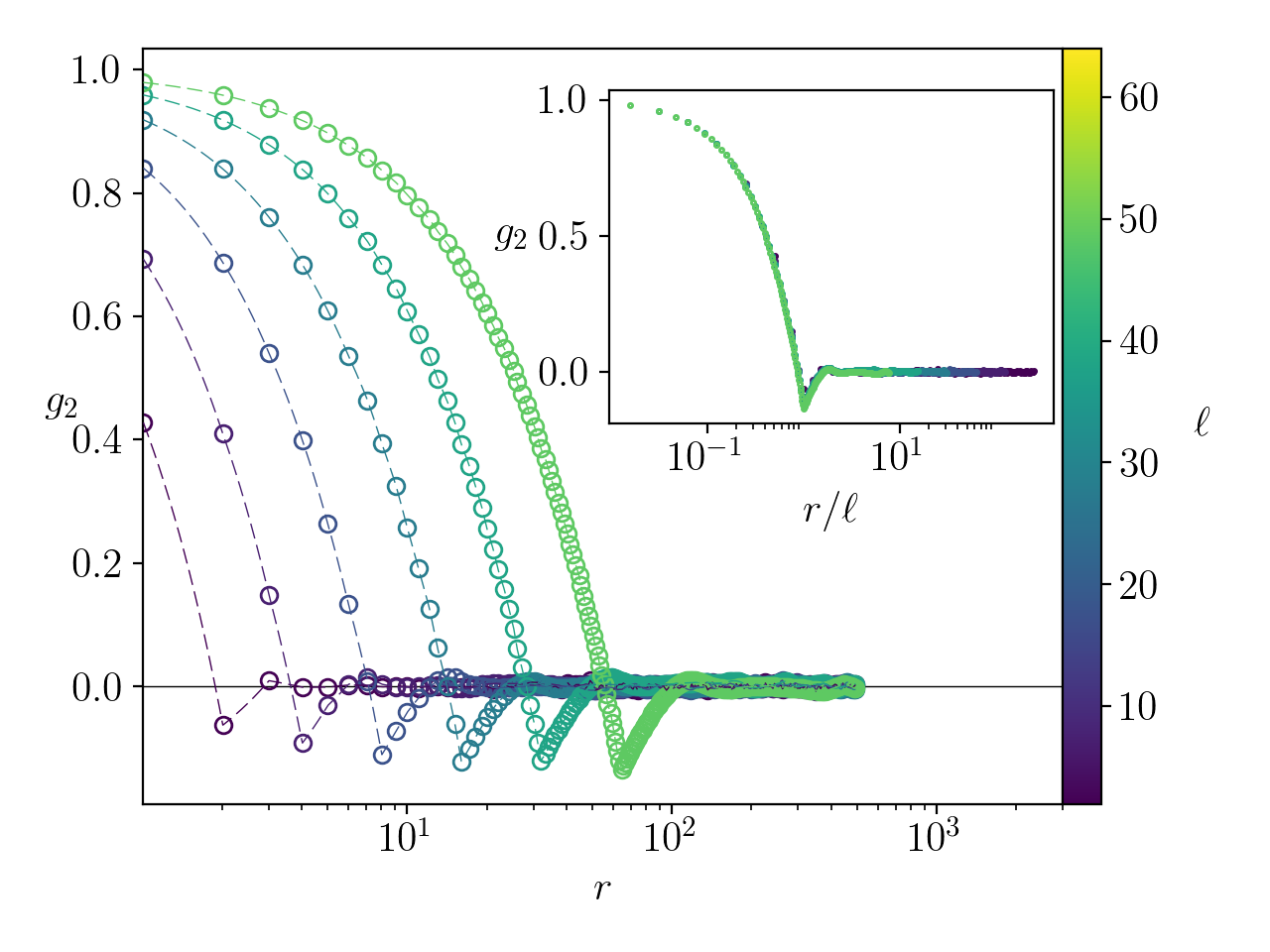}}{-0.05in}{-1.5in}
\end{subfigure}
\protect\caption{1D hard rods, with lengths $\ell=2^{i}$ ($1 \leq i \leq 6$), on a grid of length $L = 2^{16+i}$, at fixed density $\rho \equiv N_r \ell / L = 1/4$. (a) Upon decimation at intervals $\Delta > \ell$, the configurations reduce to a random sequence. The main panel shows $\xi$ as computed from $Q(\Delta)$ and from $g_2(r)$ (see panel \textit{b}) taking $\xi$ to be the value where $Q(\xi) = 0.025$ and where $g_2(\xi)$ is at its minimum. Inset shows the collapsed Q($\Delta,T$), along with the analytical solution to leading order Eq.~\ref{eq:F2} (red line). Uncollapsed curves can be seen in Fig.~1a. (b) Pair correlation function $g_2(r)$ for different rod sizes $\ell$, dashed lines are guides to the eye. Inset shows the collapsed curves.  \label{fig:1}} 
\end{figure*}

Note that ${Q}(\rho,\Delta, \ell) \rightarrow 0$ linearly with  $(\Delta - \ell)$ as $\Delta \to \ell$; this allows us to measure the correlation length numerically by finding the smallest value of $\Delta$ where ${Q}(\rho,\Delta, \ell) \approx 0$, for given $\rho$ and $\ell$.  The analytic form of  ${Q}(\rho,\Delta, \ell)$ (Eq.~\ref{eq:F2}) is shown as a universal curve as the solid line in the insert in Fig.~\ref{fig:1}a. 

\subsection{1D Ising}
\begin{figure*}
\centering
\begin{subfigure}{.5\textwidth}
\topinset{\bfseries(a)}{\includegraphics[width=1\linewidth]{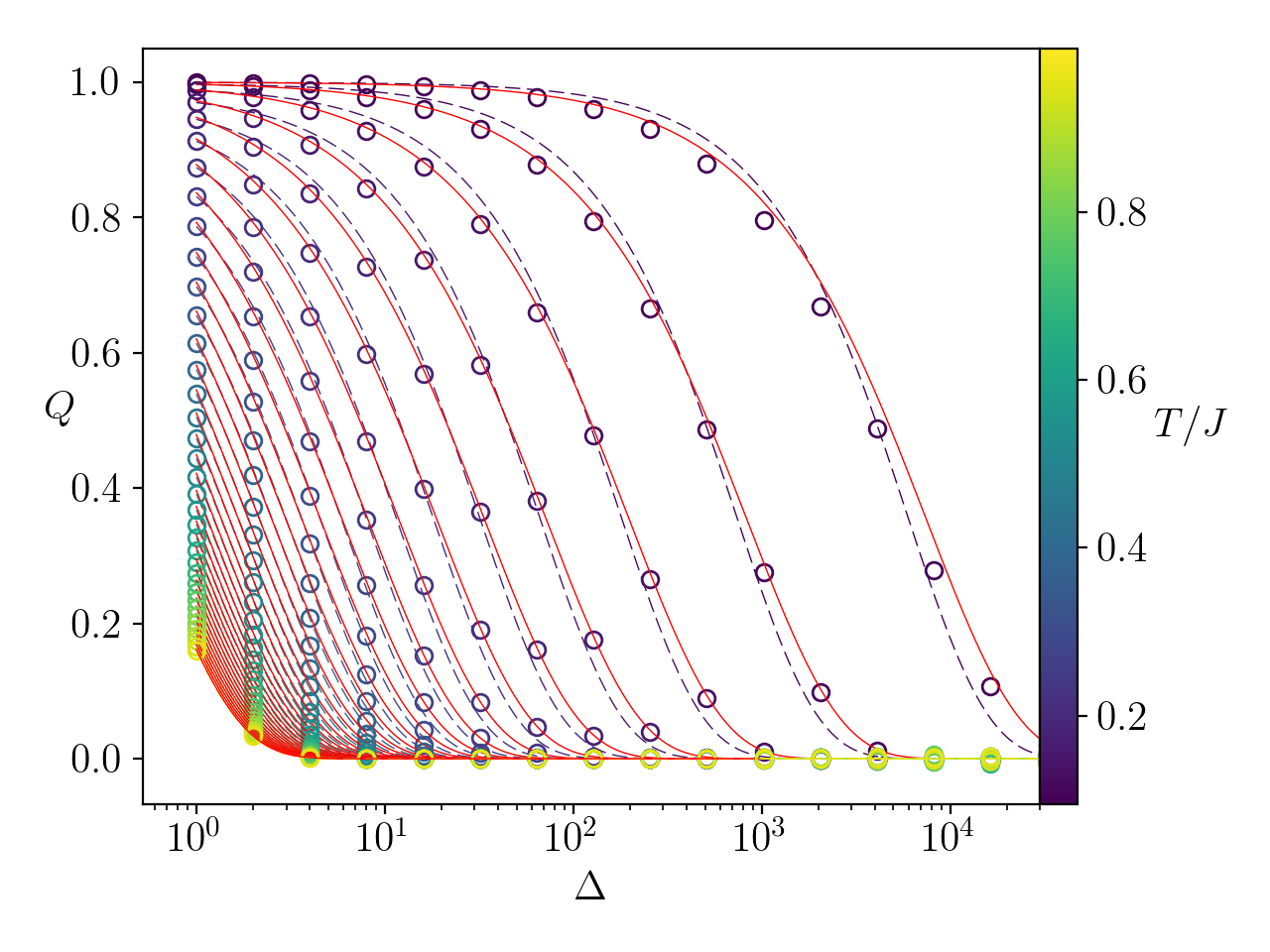}}{-0.05in}{-1.5in}
\end{subfigure}%
\begin{subfigure}{.5\textwidth}
\topinset{\bfseries(b)}{\includegraphics[width=1\linewidth]{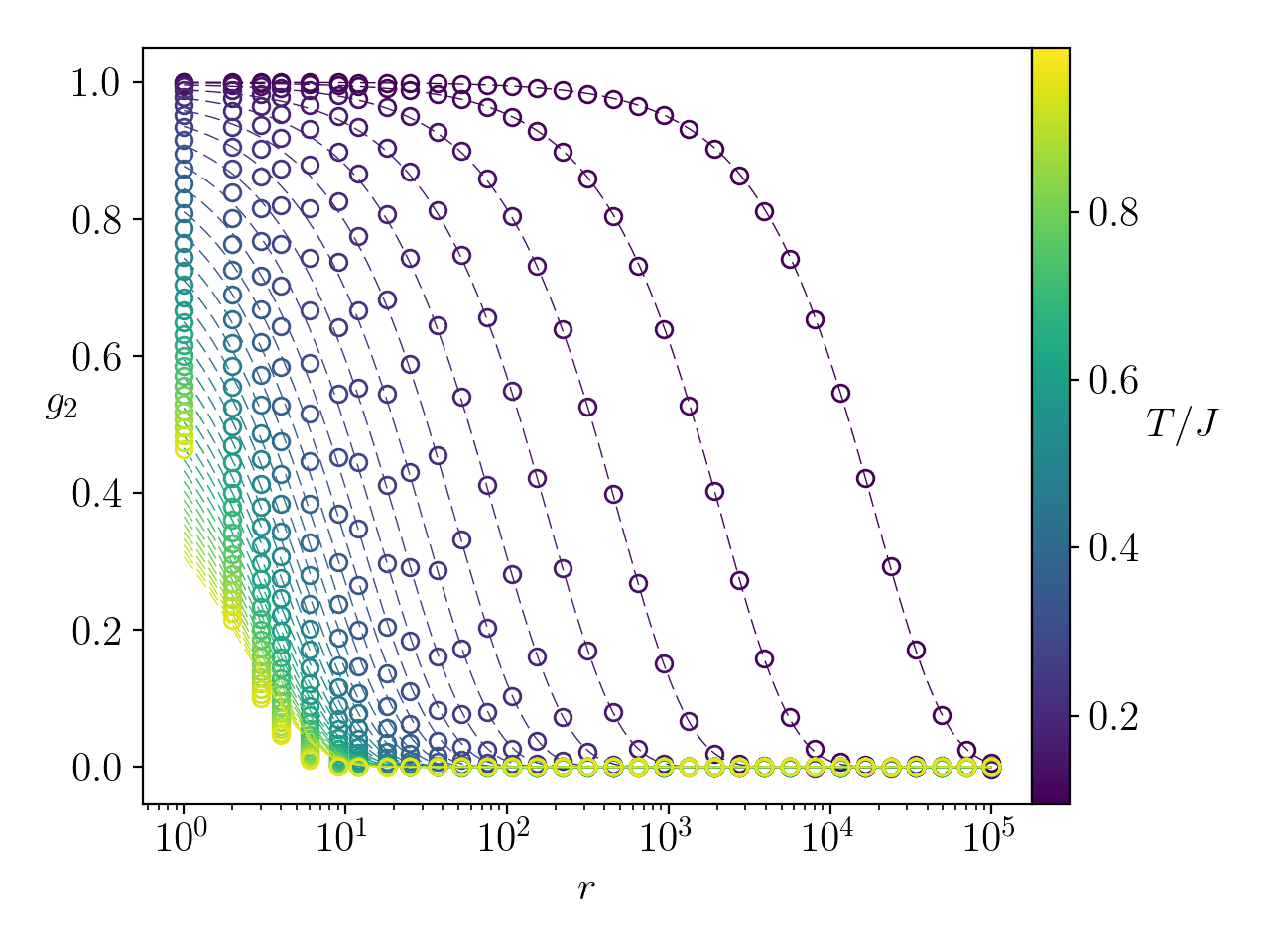}}{-0.05in}{-1.5in}
\end{subfigure}
\protect\caption{1D Ising model of size $L=2^{20}$ simulated by Wolff algorithm \cite{wolff1989collective}. (a) $Q(\Delta, T)$ for different temperatures $T$. Dashed lines are lines of best fit of the form ${Q}(\Delta) = {Q}(1)\exp(-(\Delta-1)/\xi)$ and red solid lines are from exact solution Eq.~\ref{eq:q1d}. Collapsed data are shown in the inset of Fig.~1b. (b) Pair correlation function $g_2(r)$ for several temperatures $T$. Dashed lines are lines of best fit of the form $g_2(r) = g_2(0)\exp(-r/\xi)$. Data were averaged over $200$ equilibrium configurations. For clarity we show only 1/5 of the temperatures shown in Fig.~1b of the main text.
\label{fig:2}} 
\end{figure*}

The partition function for the 1D Ising model can be computed analytically using the transfer matrix technique, in the thermodynamic limit one finds that the per-spin free energy is
\begin{equation}
f(T, B) = -\frac{1}{\beta} \log \left \{ e^{\beta J} \cosh (\beta B) + \left [ e^{2\beta J} \cosh^2(\beta B) - 2 \sinh(2\beta J) \right ]^{1/2} \right \}
\end{equation}
that in the case of zero external magnetic field ($B=0$) that we consider, simplifies to
\begin{equation}
f(T, 0) = - \frac{1}{\beta} \log \left(2 \cosh (\beta J) \right)
\end{equation}
and the per-spin entropy $S(T, 0) = -\mathrm{\partial} f(T, 0)/\mathrm{\partial} T$ is simply
\begin{equation}
S(T, 0) = \log \left ( 2 \cosh (\beta J)\right ) - \beta J \tanh (\beta J)
\label{eq:s1d}
\end{equation}
From the transfer matrix, the correlation function can be computed in a similar way, yielding
\begin{equation}
\xi(T, 0) = \frac{1}{|\log \left( \tanh (\beta J) \right) |}
\label{eq:xi1d}
\end{equation}
Solving Eq.~\ref{eq:xi1d} for $\beta J$ we find
\begin{equation}
\beta J = \tanh^{-1} (e^{-1/\xi})
\label{eq:bj1d}
\end{equation}
By substituting Eq.~\ref{eq:bj1d} into Eq.~\ref{eq:s1d},  rescaling $\xi \to \xi / \Delta$ and by noting that the per-spin entropy of a shuffled 1D Ising configuration is $S_R(\xi/\Delta) = \log 2$, we can arrive to an analytic expression for $Q \equiv (S_{R}(\xi/\Delta) - S(\xi/\Delta))/S_{R}(\xi/\Delta)$ given by using the normalized entropies instead of the CID
\begin{equation}
{Q}(\Delta, \xi) = \frac{e^{-\Delta /\xi } \coth
   ^{-1}\left(e^{\Delta /\xi }\right)+\frac{1}{2} \log
   \left(1-e^{-2
   \Delta /\xi }\right)}{\log (2)}
\label{eq:q1d}
\end{equation}
See Fig.~\ref{fig:2}a and Fig.~1b of the main text for a comparison with numerical data. Notice that a direct comparison with the numerics is only possible because for Ising 1D decimation by $\Delta$ yields configurations with a correlation length $\xi/\Delta$ that are valid Ising configurations whose temperature is given by Eq.~\ref{eq:bj1d}.

\subsection{2D Ising}
The partition function for the zero field 2D Ising model was solved analytically by Onsager \cite{onsager1944crystal} and can be expressed in terms of a definite-integral, so that the per-spin free energy takes the form

\begin{equation}
\beta f(T, 0) = - \log 2 - \frac{1}{2 \pi^2} \iint_0^{2\pi} \log P(T, \mathbf{k}) \mathrm{d} \mathbf{k} 
\end{equation}
where
\begin{equation}
P(T, \mathbf{k}) = \cosh^2(2 \beta J) - \sinh(2 \beta J) (\cos k_1 + \cos k_2)
\end{equation}
and the per-spin entropy $S(T, 0) = -\mathrm{\partial} f(T, 0)/\mathrm{\partial} T$ is simply
\begin{equation}
S(T, 0) = \log 2 + \frac{1}{2\pi^2} \iint_0^{2\pi}  \log P(T, \mathbf{k}) \mathrm{d} \mathbf{k}  + \frac{\beta J}{\pi^2} \iint_0^{2 \pi} \frac{\cosh(2 \beta J)(\cos k_1 + \cos k_2 - 2 \sinh(2 \beta J))}{P(T, \mathbf{k})} \mathrm{d} \mathbf{k}
\label{eq:s2d}
\end{equation}
The correlation function can also be computed, yielding the correlation length in closed form \cite{codello2015approximating}
\begin{equation}
\xi(T, 0) = \frac{1}{\log \frac{\sqrt{2} - 1}{\frac{1 - e^{-2\beta J}}{1 + e^{-2\beta J}}}}
\label{eq:xi2d}
\end{equation}
Solving Eq.~\ref{eq:xi2d} for $\beta J$ we find
\begin{equation}
\beta J = \frac{1}{2} \log \frac{e^{1 /\xi} -1 + \sqrt{2}}{e^{1 /\xi} + 1 - \sqrt{2}}
\label{eq:bj2d}
\end{equation}
so that
\begin{equation}
\begin{aligned}
\cosh(2\beta J) &= \frac{e^{2/\xi} + 3 - 2 \sqrt{2}}{e^{2/\xi} -3 + 2 \sqrt{2}}\\
\sinh(2 \beta J) &= \frac{2(\sqrt{2} - 1) e^{1/\xi}}{2^{2/\xi}-3 + 2\sqrt{2}}
\end{aligned}
\label{eq:trig}
\end{equation}

By substituting Eqs.~\ref{eq:bj2d}-\ref{eq:trig} into Eq.~\ref{eq:s2d},  rescaling $\xi \to \xi / \Delta$ and by noting that for $T > T_c$ the per-spin entropy of a shuffled 2D Ising configuration is $S_R(\xi/\Delta) = \log 2$, we can arrive to an analytic expression for $Q_{T>T_c} \equiv (S_{R}(\xi/\Delta) - S(\xi/\Delta))/S_{R}(\xi/\Delta)$ (given by using the normalized entropies instead of the CID) that can be readily evaluated.

\begin{figure*}
\centering
\begin{subfigure}{.5\textwidth}
\topinset{\bfseries(a)}{\includegraphics[width=1\linewidth]{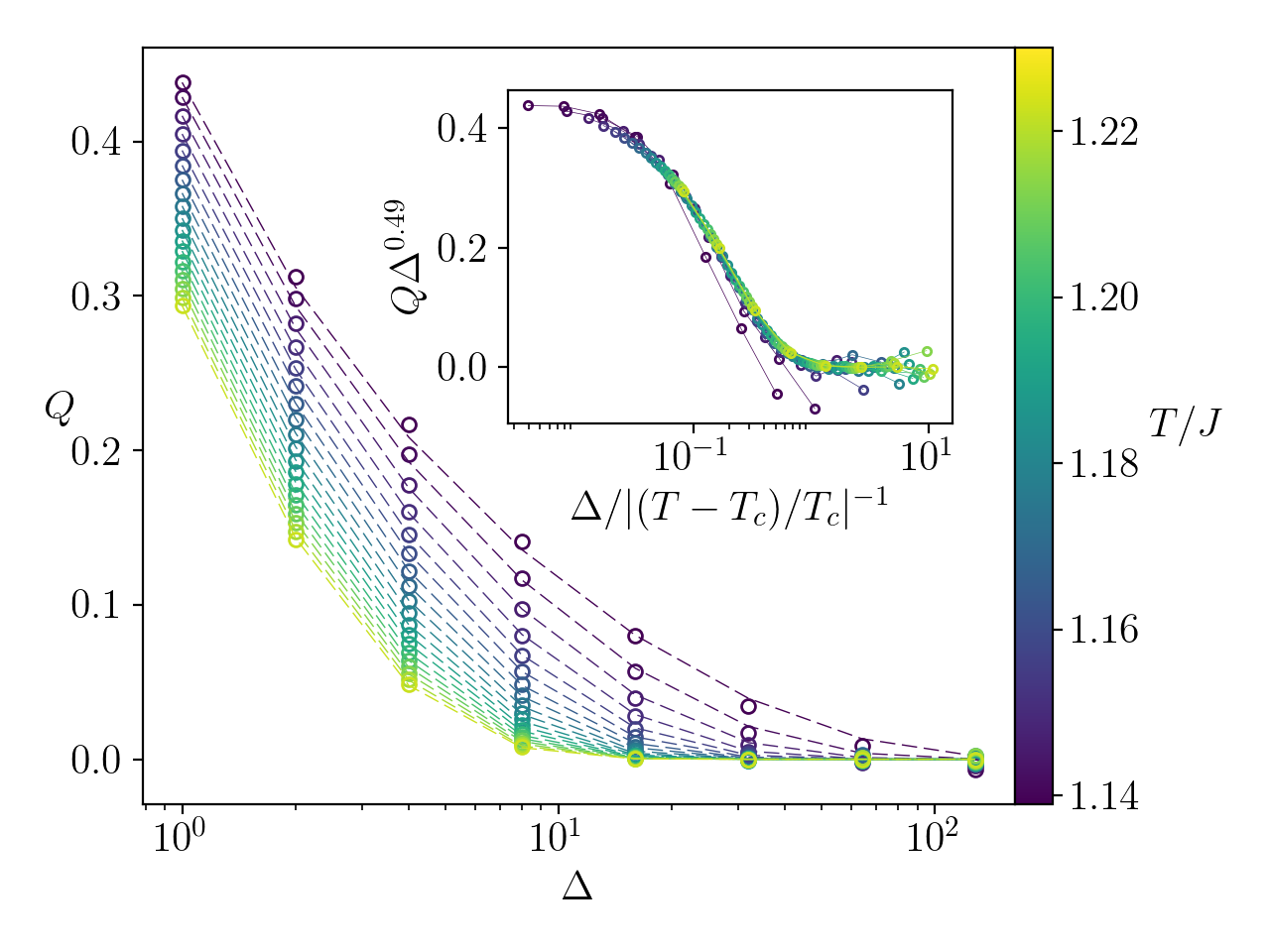}}{-0.05in}{-1.5in}
\end{subfigure}%
\begin{subfigure}{.5\textwidth}
\topinset{\bfseries(b)}{\includegraphics[width=1\linewidth]{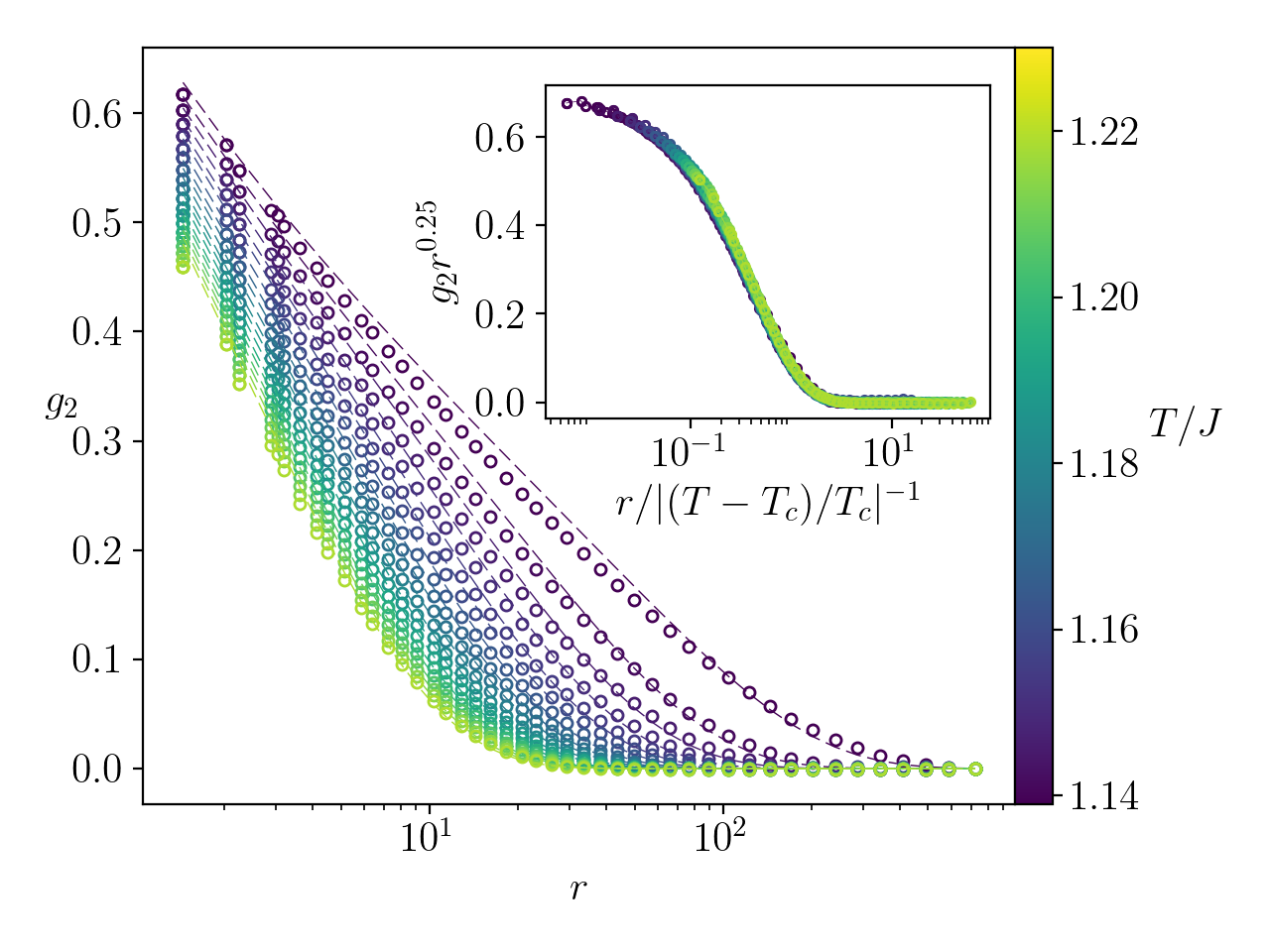}}{-0.05in}{-1.5in}
\end{subfigure}
\protect\caption{2D Ising model (size $L=2^{10} \times 2^{10}$) simulated by Wolff algorithm \cite{wolff1989collective}. (a) $Q(\Delta, T)$ for different temperatures $T>T_c$ with $T_c \approx 1.1345$. Dashed lines are lines of best fit of the form ${Q}(\Delta) = {Q}(1)\exp(-(\Delta-1)/\xi)/\Delta^\theta$. Fits were performed iteratively optimizing $\theta$ over all temperatures and then $\xi$ independently for each temperature until convergence, yielding $\theta\approx1/2$. Inset shows curves scaled by the theoretical scaling $\xi \sim |T-T_c|/T_c$. (b) Pair correlation function $g_2(r)$ for several temperatures $T>T_c$. Dashed lines are lines of best fit of the form $g_2(r) = g_2(0)\exp(-r/\xi)/r^\eta$. Fits were performed as in \textit{a} yielding $\eta=0.28$ (cf. exact result $\eta=0.25$). Inset shows curves scaled by the theoretical scaling $\xi \sim |T-T_c|/T_c$. Data were averaged over $200$ equilibrium configurations. For clarity we show only 1/5 of the temperatures shown in Fig.~2 of the main text and in Fig.~\ref{fig:5}. \label{fig:3}} 
\end{figure*}

\begin{figure*}
\centering
\begin{subfigure}{.7\textwidth}
\includegraphics[width=1\linewidth]{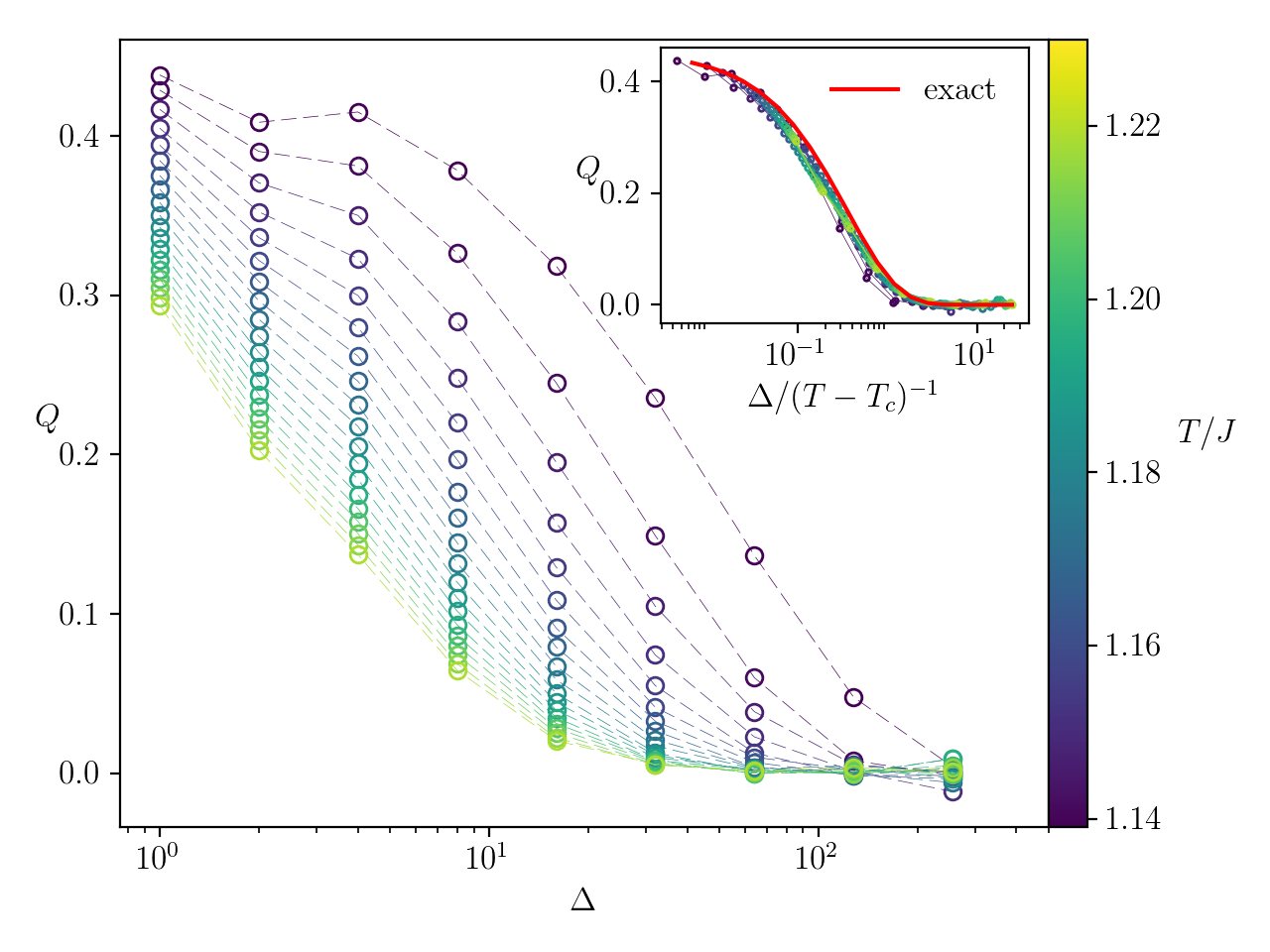}
\end{subfigure}%
\protect\caption{2D Ising model (size $L=2^{10} \times 2^{10}$) simulated by Wolff algorithm \cite{wolff1989collective}. (a) $Q(\Delta, T)$ computed by \textit{majority rule} for different temperatures $T>T_c$ with $T_c \approx 1.1345$. Dashed lines are a guide to the eye. Inset shows curves scaled by the theoretical scaling $\xi \sim |T-T_c|/T_c$ and red line is from the analytical solution from Eqs.~\ref{eq:s2d}-\ref{eq:trig} but adjusted by a constant multiplicative factor ($\approx 0.8$), see text for a discussion. Data were averaged over 200 equilibrium configurations. For clarity we show only 1/5 of the temperatures considered in Fig.~2 of the main text. \label{fig:4}}
\end{figure*}

Unlike for the 1D Ising model we cannot perform a direct comparison of the thus computed exact expression for $Q_{T>T_c}(\Delta, \xi)$ with that obtained from decimation in Fig.~2 and Fig.~\ref{fig:3}. The reason for this is that while decimating by $\Delta$ correctly yields configurations with correlation length $\xi/\Delta$, these configurations are not equilibrium Ising configurations whose temperature would be given by Eq.~\ref{eq:bj2d}. This can be seen for instance from the fact that magnetization ($m$) is invariant under decimation. For a transformation to yield a valid equilibrium configuration, the magnetization must approach $m=1$ when starting below $T_c$, and for a finite-size system it must approach $m=0$ when starting above $T_c$ (although in the thermodynamic limit $m=0$ for all $T \geq T_c$). A transformation for which one observes a change in magnetization upon changing the correlation length is Kadanoff's blocking transformation known as \textit{majority rule} \cite{kadanoff1966scaling}. Just like for decimation, we take blocks of size $\Delta^d$ but instead of choosing the upper-left spin, majority rule yields spin $+1$ if $m_{\text{block}} > 0$, $-1$ if $m_{\text{block}} < 0$ and samples $\{+1, -1\}$ with equal probability when $m_{\text{block}}=0$ \cite{binney1992theory}. We compare the exact expression for $Q_{T>T_c}(\Delta, T)$ with results obtain from CID using majority rule in the inset of Fig.~\ref{fig:4} and find good agreement, up to a constant multiplicative factor ($\approx 0.8$) in $Q$ because for a finite-size system the CID computed by LZ77 upper bounds the entropy of the 2D Ising model \cite{martiniani2019quantifying}.

\section{Rudin-Shapiro sequence}

The (Golay-)Rudin-Shapiro sequence (RS) \cite{golay1949multi, golay1951static, shapiro1952extremal, rudin1959some} is a quasiperiodic binary sequence that can be generated using the recursion relations $\{A \to AB, B \to AD, C \to CD, D \to CB\}$, and then making the replacements $A\to11,B\to1{\bar 1},C\to{\bar 1}{\bar 1}, D\to{\bar 1}1$. The  CID of the RS varies with system size $N$ and approaches $0$ as $\log_2(N)^2/N$ \cite{constantinescu2007lempel, martiniani2019quantifying}, despite having a flat Fourier transform (like a random sequence, see Fig.~4a) and no two-point correlations.

\section{Supplementary data}

\begin{figure*}
\centering
\begin{subfigure}{.5\textwidth}
\topinset{\bfseries(a)}{\includegraphics[width=1\linewidth]{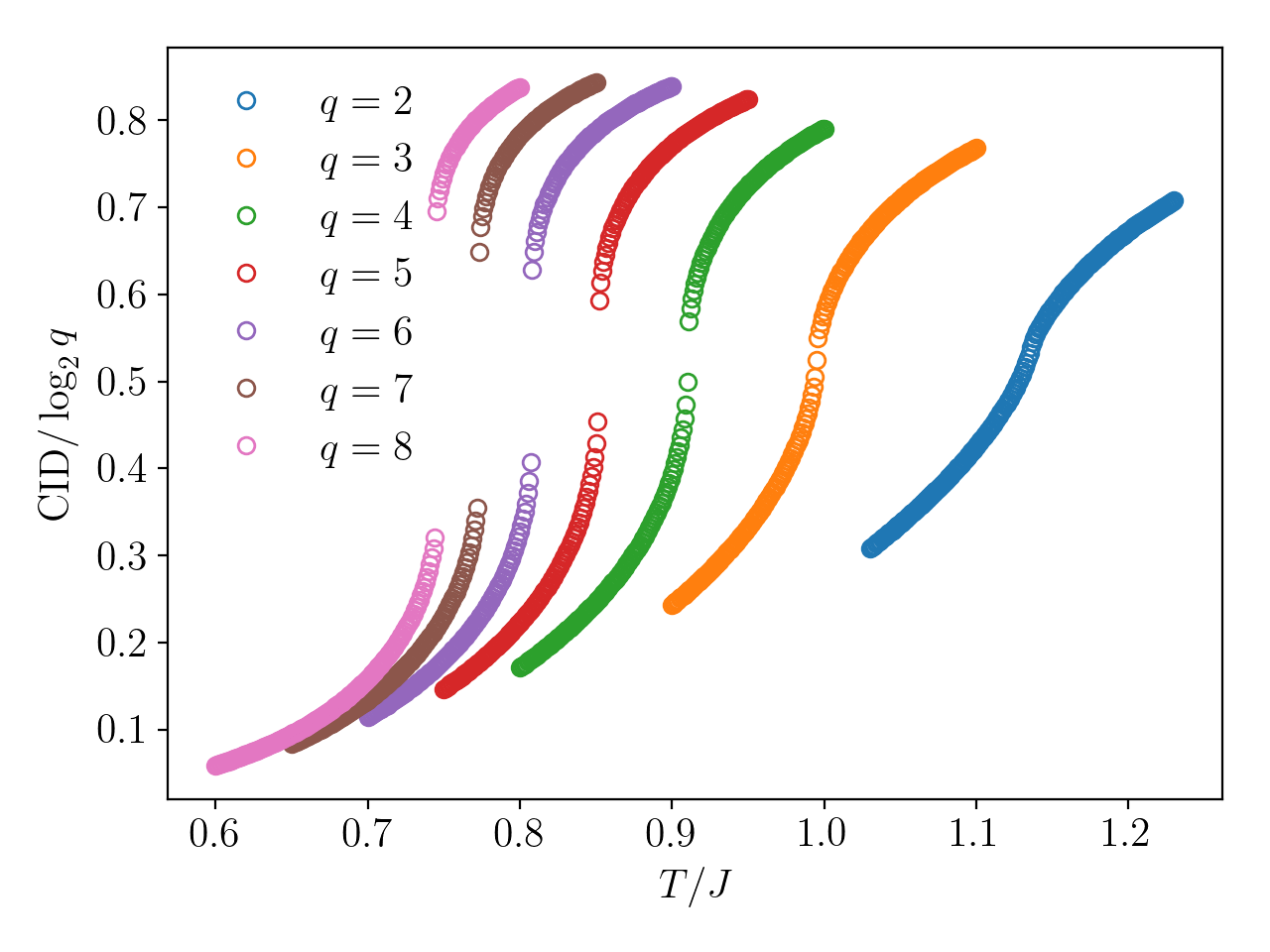}}{-0.05in}{-1.5in}
\end{subfigure}%
\begin{subfigure}{.5\textwidth}
\topinset{\bfseries(b)}{\includegraphics[width=1\linewidth]{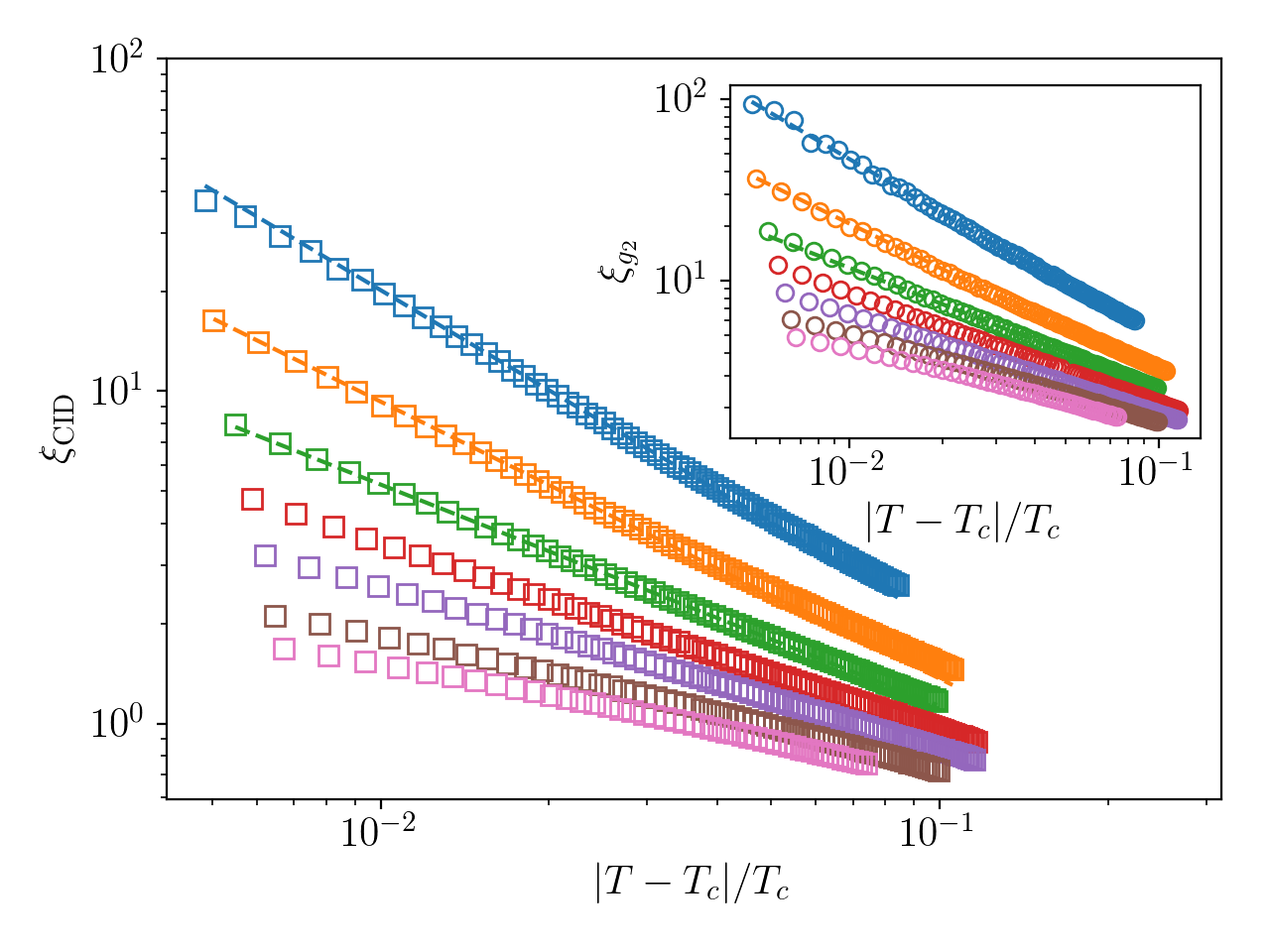}}{-0.05in}{-1.5in}
\end{subfigure}
\protect\caption{2D \textit{q}-state Potts model (size $L=2^{10} \times 2^{10}$, $2\leq q \leq 8$) simulated by Wolff algorithm \cite{wolff1989collective}. (a) CID shows that for $q>3$ phase transitions increasingly acquire discontinuous character (they are discontinuous for $q>4$). We identify transition temperatures from $\text{arg} \max \partial \mathrm{CID} / \partial T$ yielding $T_c \approx [1.1345, 0.995 , 0.910, 0.851 , 0.807 , 0.773 , 0.745 ]$. (b) Correlation lengths $\xi$ extracted by fitting $Q(\Delta)  = Q(1)\exp(-(\Delta - 1)/\xi)/\Delta^\theta$. Inset shows the correlation lengths extracted by fitting the pair-correlation function $g_2(r) = g_2(0)\exp(-r/\xi)/r^\eta$. Fits were performed iteratively optimizing $\eta$ over all temperatures and then $\xi$ independently for each temperature until convergence, yielding $\theta \approx [0.48, 0.53, 0.53, 0.52, 0.51, 0.45, 0.42]$ and $\eta \approx [0.28, 0.31, 0.34, 0.37, 0.41, 0.44, 0.47]$. Dashed lines show the theoretical scaling $\xi \sim |T-T_c|^{-\nu}$ with $\nu = [1, 5/6, 2/3]$. For $q>3$ we observe increasingly large deviations from linearity. Data were averaged over 200 equilibrium configuration. \label{fig:5}} 
\end{figure*}

\begin{figure*}
\centering
\begin{subfigure}{.5\textwidth}
\topinset{\bfseries(a)}{\includegraphics[width=1\linewidth]{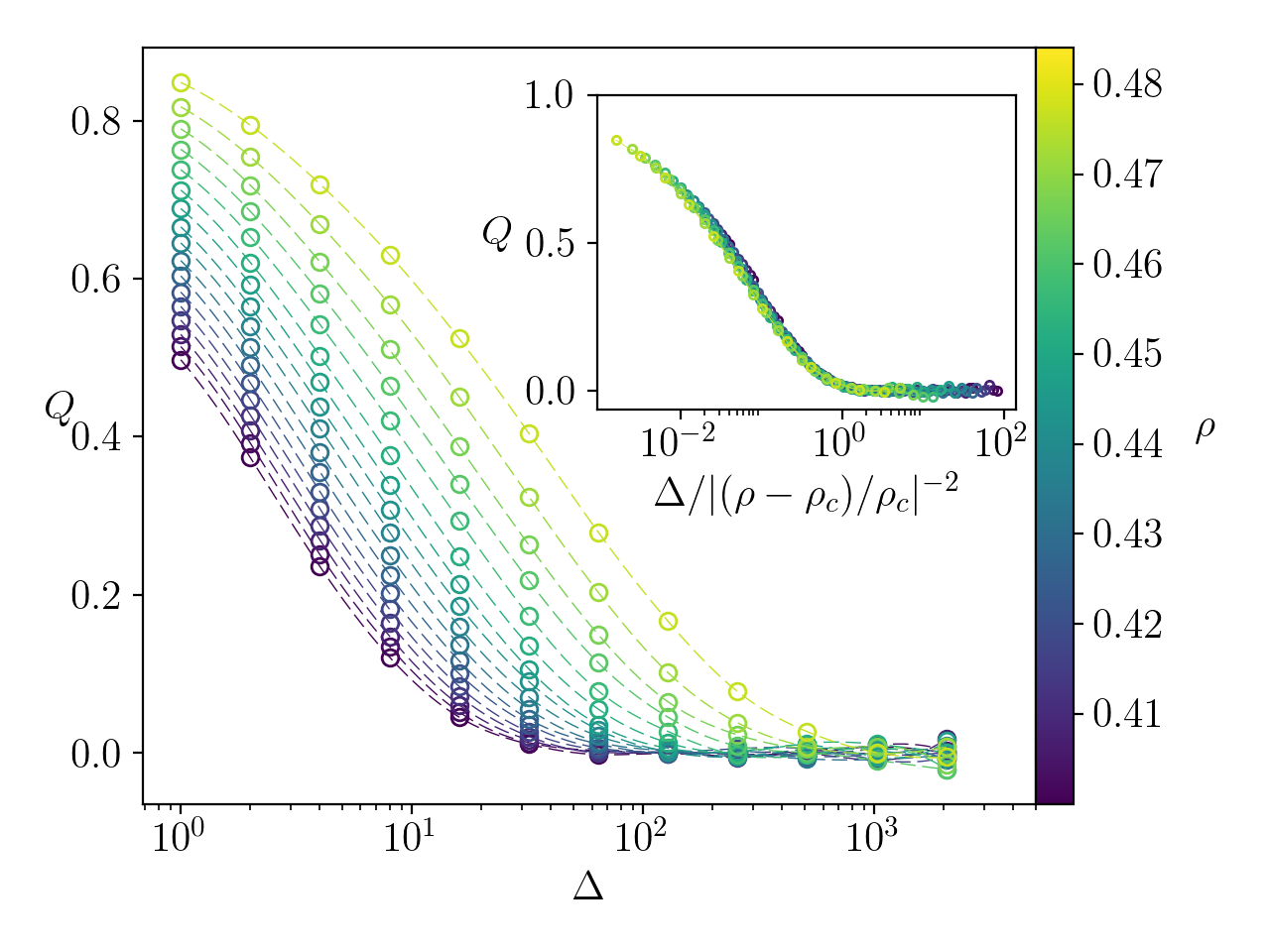}}{-0.05in}{-1.5in}
\end{subfigure}%
\begin{subfigure}{.5\textwidth}
\topinset{\bfseries(b)}{\includegraphics[width=1\linewidth]{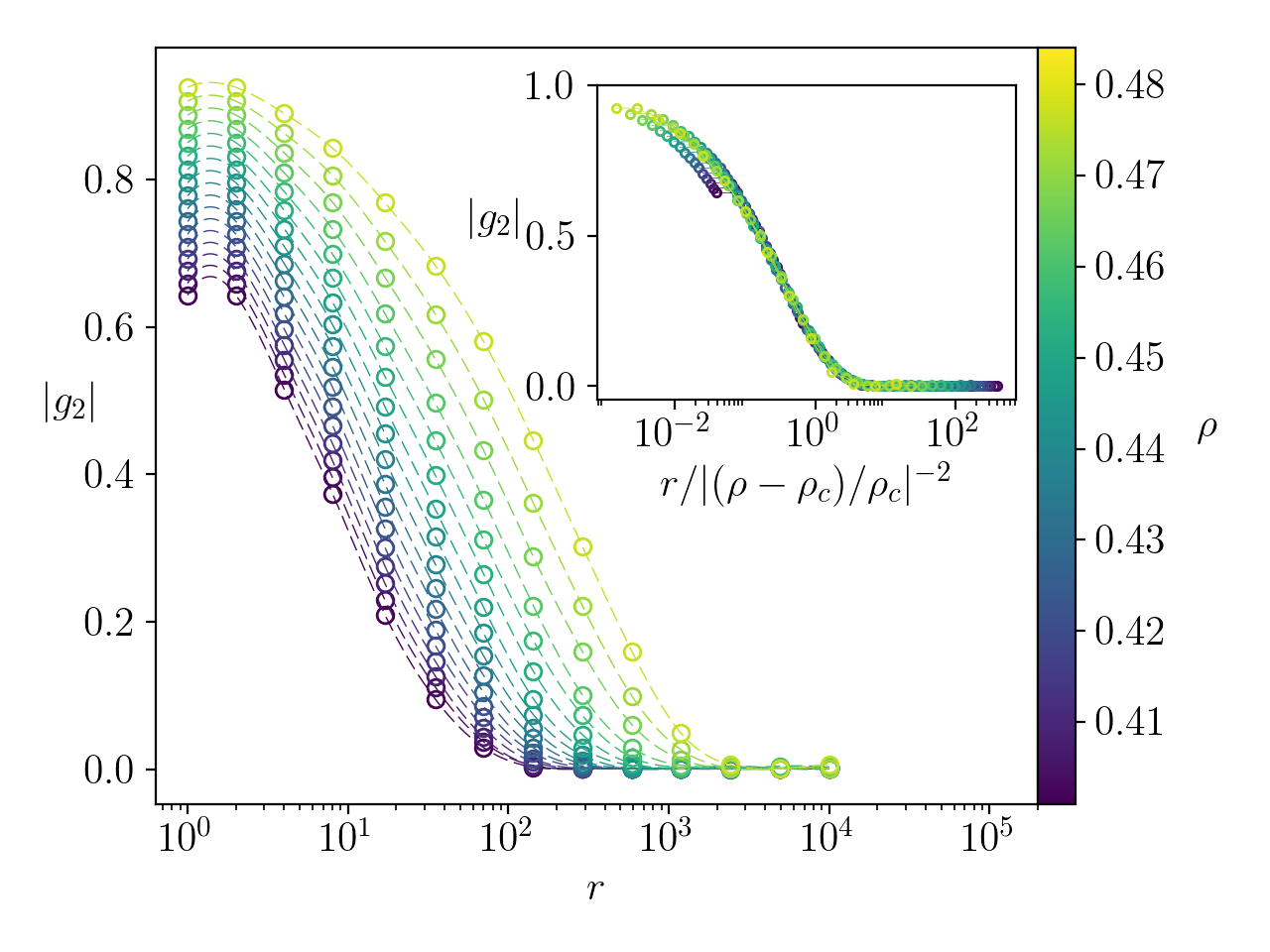}}{-0.05in}{-1.5in}
\end{subfigure}
\protect\caption{1D  Conserved Lattice Gas model of size $L=2^{17}$ simulated starting from randomly sampled states. (a) $Q(\Delta, \rho)$ for several densities $\rho<\rho_c$ with $\rho_c=0.5$. Dashed lines are guide to the eye. Inset shows data collapsed by fitted scaling $\xi \sim |\rho-\rho_c|^2$. (b) Pair correlation function $g_2(r)$ for several densities $\rho < \rho_c$. Dashed lines are guide to the eye. Inset shows data collapsed by fitted scaling $\xi \sim |\rho-\rho_c|^2$. Data were averaged over 15 independently generated configurations. For clarity we show only 1/5 of the temperatures shown in Fig.~3 of the main text and in Fig.~\ref{fig:7}. \label{fig:6}} 
\end{figure*}

\begin{figure*}
\centering
\includegraphics[width=0.7\linewidth]{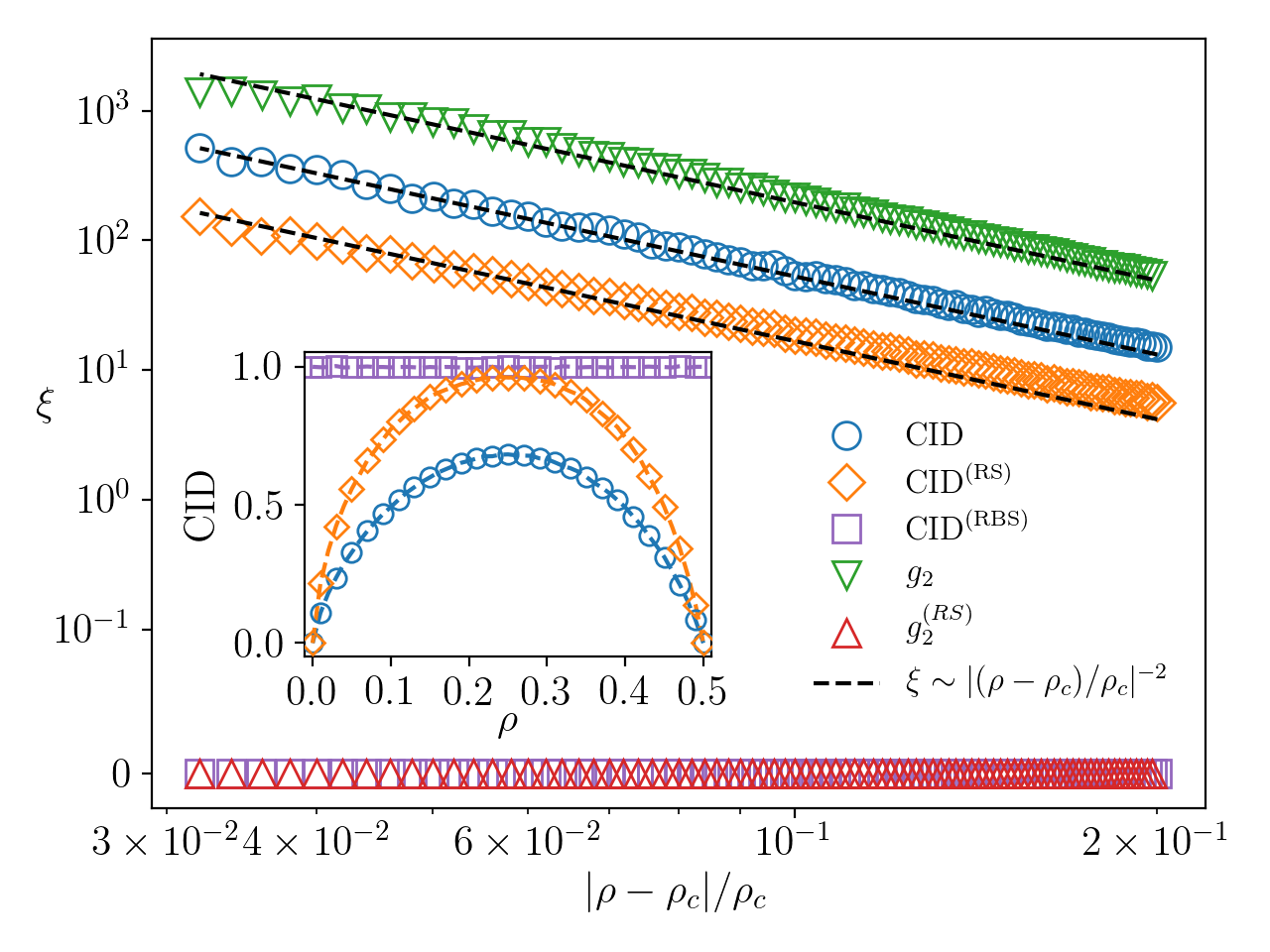}
\protect\caption{Correlation lengths $\xi$ for the 1D Conserved Lattice Gas model of size $L=2^{17}$ starting from randomly sampled states, extracted from $Q(\Delta)$ and $g_2(r)$. Results are shown for the unaltered system, as well as for the Rudin-Shapiro, RS-cloaked, and Random Bernoulli Sequence (RBS) randomly-cloaked sequences. Correlation lengths are extracted taking $\xi$ to be the value where $Q(\xi) = 0.05$ and $g_2(\xi) = 0.05$. The black dashed lines show the fitted scaling $\xi \sim |(\rho - \rho_c)/\rho_c|^{-2}$, where $\rho_c = 0.5$. Data were averaged over 15 independently generated configurations. Notice how CID correctly identifies the correlation length scaling of the sequence cloaked by RS while $g_2(r)$ does not. Inset: CID as a function of $\rho$ for the uncloaked and cloaked systems. The RS-cloaked system has nontrivial CID, but the randomly-cloaked system does not. \label{fig:7}} 
\end{figure*}

In addition to the figures already discussed in this Supplementary Information, in Fig.~\ref{fig:5} we show results for the CID and the correlation length $\xi$ extracted from $Q(\Delta)$ and $g_2(r)$ for 2D \textit{q}-state Potts models with $2 \leq q \leq 8$. In Fig.~\ref{fig:6} we show $Q(\Delta, \rho)$ and $g_2(r, \rho)$ for the 1D CLG. Insets show the same data collapsed by the fitted scaling $\xi \sim |\rho-\rho_c|^2$. In Fig.~\ref{fig:7} (inset) we show the CID for 1D CLG configurations for several densities $\rho < \rho_c$, and for the same configurations cloaked by Rudin-Shapiro (RS) and by Random Bernoulli Sequence (RBS). RBS-cloaked CLG has a flat $\text{CID} = 1$ indicating a correlation-less, maximally disordered system, but RS-cloaked CLG retains much of its correlations. The main panel in Fig.~\ref{fig:7} shows the correlation length extracted from $Q(\Delta)$ and $g_2(r)$ for the same CLG configurations (cloaked and uncloaked).

\section{Implementation details}

CID was computed, as in Ref.~\cite{martiniani2019quantifying}, by the unrestricted Lempel Ziv string-matching algorithm \cite{ziv1977universal, shields1999performance}, also known as LZ77, using the open-source library \textit{Sweetsourcod} \cite{sweetsourcod} that wraps the linear-time algorithms for LZ77 factorization by Karkkainen, Kempa and Puglisi \cite{karkkainen2013linear}. We adopt the KKP2 algorithm available at Ref.~\cite{lz77alg}, capable of performing the LZ77 factorization in $O(N)$ time complexity. 2D configurations were serialized prior to compression using a Hilbert scan, as described in Ref.~\cite{martiniani2019quantifying}. The implementation of the Hilbert curve in \textit{Sweetsourcod} is based on the method by Skilling \cite{skilling2004programming} and adapted from Ref.~\cite{hilbertcurve}.

\bibliography{bibliography}